\documentclass[pra,aps,showpacs]{revtex4}
\usepackage{bm}
\usepackage{graphics}
\usepackage[dvips]{color}
\usepackage{dcolumn}% Align table columns on decimal point
\newcounter{six}
\newcounter{sixteen}
\begin{document}
\title{Laser probing of single-particle energy gap of a Bose gas in an optical lattice in the Mott insulator phase}
\author{S. Konabe$^1$, T. Nikuni$^1$, and M. Nakamura$^2$}
\affiliation{$^1$Department of Physics, Faculty of Science, Tokyo University of Science, 
1-3 Kagurazaka, Shinjuku-ku, Tokyo, Japan, 162-8601\\
$^2$Department of Applied Physics, Faculty of Science, Tokyo University of Science, 
1-3 Kagurazaka, Shinjuku-ku, Tokyo, Japan, 162-8601}
\date{\today}

\begin{abstract}
We study single-particle excitations in the Mott insulator phase of a Bose gas in an optical lattice.
The characteristic feature of the single-particle spectrum in the Mott insulator phase is the existence of an energy gap between the particle and hole excitations.
We show that this energy gap can be directly probed by an output coupling experiment.
We apply the general expression for the output current derived by Luxat and Griffin, which is given in  terms of the single-particle Green's functions of a trapped Bose gas, to the Mott insulator phase using the Bose-Hubbard model.
We find that the energy spectrum of the momentum-resolved output current exhibits two characteristic peaks corresponding to the particle and hole excitations, and thus it can be used to detect the transition point from the Mott insulator to superfluid phase where the energy gap disappears.  
\end{abstract}

\maketitle
\section{introduction}\label{sec:intro}
Ultracold atomic gases in optical lattices provide a new framework for the experimental 
study of many-body quantum systems by making use of their remarkable degree of experimental control of the relevant parameters. 
In particular a recent pioneering experiment by Greiner \textit{et al.}~\cite{greiner} observed a phase transition from the superfluid to Mott insulator phase in a  ultracold gas of $^{87}$Rb atoms trapped in a three-dimensional simple-cubic optical potential.

In the experiment of Ref.~\cite{greiner}, the characteristic feature of the superfluid phase appeared in the interference pattern in the absorption image of the atomic cloud after ballistic expansion, which is due to the coherence of the Bose condensed atoms. 
The phase transition to the Mott insulator phase was signaled by the disappearance of the interference pattern. 
However, it is not obvious how one can precisely identify the transition point from the interference visibility. 
Moreover, recent papers by Roth and Burnett~\cite{roth1,roth2} argue that the interference pattern contains only the information on the quasi-momentum distribution of the lattice system, but has no direct information on the superfluid fraction. 
Therefore the disappearance of the interference pattern does not provide direct proof of the Mott transition~\cite{roth1,roth2,kashurnikov}.
On the other hand, in the Mott insulator phase, their characteristic feature appears in its single-particle excitation spectra, which have a gap between particle and hole excitations.
The experiment by Greiner \textit{et al.}~\cite{greiner} also provided the spectroscopic study of this energy gap by applying a potential gradient.

In this paper, we propose to use an output coupling current as a more direct probe for detecting the Mott gap of a Bose gas in an optical lattice.
Recent papers~\cite{japha,choi,luxat} discussed the output coupling current from the trapped atoms at finite 
temperature, and pointed out that the analysis of the output current serves as a probe of many-body states of a trapped gas.
In particular, Luxat and Griffin~\cite{luxat} developed a general formalism of the output coupling current by 
using the linear response theory. 
Their formalism is useful because one only has to concentrate on calculating single-particle correlation functions of the system of interest. 
In this paper, we calculate the output coupling current from a Bose gas in an optical lattice in the Mott insulator phase, 
and explicitly show how the characteristic feature of single-particle excitations appears in the energy spectrum of the output-coupling current.

To describe a Bose gas in an optical lattice, we use a Bose-Hubbard model~\cite{fisher,jaksch}.
This model has been shown to exhibit the phase transition from the superfluid to the Mott insulator at zero-temperature, 
and its phase diagram has been investigated by many authors both analytically and numerically~\cite{fisher,sheshadri,oosten,elstner,kuhner,batrouni,freericks}.
The mean-field theory predicts the transition point, at which 
the order parameter for the superfluid states vanishes, $U/zt=5.83\equiv U_{\rm c}/zt$~\cite{fisher,sheshadri,oosten}, where $U$ is an on-site interaction, $t$ is a hopping parameter,  and $z$ is the number of the nearest sites in a Bose-Hubbard model. 
More recently, finite-temperature behaviors of a Bose gas in an optical lattice have been studied in Refs.~\cite{dickerscheid,buonsante}.
In Sec.~\ref{sec:bose}, we briefly discuss the mean-field phase diagram of a Bose-Hubbard model at finite temperatures using the Landau theory of the phase transition. 
A detailed microscopic derivation of the Landau free energy of a Bose-Hubbard model is presented in Appendix 
\ref{sec:derivation}.

In Sec.~\ref{sec:correlation}, we use the formalism of Luxat and Griffin~\cite{luxat} to express the output coupling current from an optical lattice in terms of the single-particle 
correlation functions of the Bose-Hubbard model. 
In Sec.~\ref{sec:many} we then calculate the single-particle Green's function of the Bose-Hubbard model 
by using the standard basis operator formalism developed in \cite{haley}. 
This formalism is useful for calculating excitations in the strong-coupling ($U>t$) regime near the Mott transition. 
We solve the equation of motion for the single-particle Green's function in the Mott insulator phase by the random-phase decoupling approximation. 
The poles of this Green's function provide dispersion relations of particle and hole excitations. 
These dispersion relations exhibit an energy gap, which vanishes at a critical point and grows with increasing on-site interaction as well as with increasing temperature. 
We will show that this behavior of the energy gap can be directly probed by the output coupling current.

In Sec.~\ref{sec:output}, we give an analytical expression for the output coupling current from an optical lattice in 
terms of the spectrul functions of the Bose-Hubbard model calculated in Sec.~\ref{sec:many}. 
We then explicitly calculate the output current as a function of the energy transfer from probe laser field, and show that a momentum-resolved output coupling current exhibits two characteristic peaks corresponding to the particle and hole 
excitations, allowing for direct observation of the Mott gap. 
Thus, one can use a momentum-resolved output-coupling current as a direct probe of the phase transition from a Mott insulator phase to superfluid phase.

\section{Mean-field phase diagram of a Bose-Hubbard model}\label{sec:bose}
In this section, we discuss the superfluid-Mott insulator transition of a Bose gas trapped in an optical lattice potential. 
Our system is described by the Hamiltonian
\begin{eqnarray}\label{mean:optical}
\mathcal{H}_{{\rm t}}&=&\int{\textit d}{\bf r}\ \psi_{{\rm t}}^{\dag}({\bf r})
\left[-\frac{\hbar^2}{2m}\nabla^2+V_{{\rm op}}({\bf r})+V_{{\rm t}}({\bf r})-\mu\right]\psi_{\rm t}({\bf r})\nonumber\\
& &{}+\frac{1}{2}\frac{4\pi a_s\hbar^2}{m}
\int{\it d}{\bf r}\ \psi_{\rm t}^{\dag}({\bf r})\psi_{\rm t}^{\dag}({\bf r})\psi_{\rm t}({\bf r})\psi_{\rm t}({\bf r}),
\end{eqnarray}
where $\psi_{\rm t}({\bf r})$ is the Bose field operator for a trapped atomic state. 
Here $V_{\rm t}({\bm r})$ is the magnetic trap potential and $V_{\rm op}({\bf r})=V_0\sum_{j=1}^3\sin^2(2\pi x_j/\lambda)$ is the simple-cubic optical lattice potential with the lattice constant $a=\lambda/2$. 
The interatomic interaction is approximated by the short-range potential with the $s$-wave scattering length $a_{\rm s}$. 
We assume that the band gap energy between the first and the second excitation band is large 
compared to the temperature, and thus only the first band is thermally occupied. 
The depth of the optical lattice wells is assumed large enough to make the atomic wave functions well localized on the 
individual lattice sites. 
Under these assumptions, one can expand the field operator $\psi_{\rm t}({\bf r})$ of the trapped atoms in terms of the Wannier function $w({\bf r})$
\begin{eqnarray} 
\psi_{\rm t}({\bf r})=\sum_ib_iw({\bf r}-{\bf r}_i),
\end{eqnarray} 
and rewrite $\mathcal{H}_{\rm t}$ into the Bose-Hubbard model
\begin{equation}\label{bose-hubbard}
\mathcal{H}_{\rm t}
=-t\sum_{<ij>}b_i^{\dag}b_j+\frac{U}{2}\sum_ib_i^{\dag}b_i^{\dag}b_ib_i
-\sum_i\mu_ib_i^{\dag}b_i,
\end{equation}
where $b_i^{\dag}$ and $b_i$ are the Bose creation and annihilation operators of atoms at the lattice site $i$, respectively. 
The parameters in Eq.(\ref{bose-hubbard}) are defined by
\begin{eqnarray}
t&=&\int{\it d}{\bf r}\ w^*({\bf r}-{\bf r}_i)\left[
-\frac{\hbar^2}{2m}\nabla^2+V_{\rm op}({\bf r})\right]w({\bf r}-{\bf r}_j),\\
U&=&\frac{4\pi a_s\hbar^2}{m}\int{\it d}{\bf r}\ |w({\bf r})|^4,\\
\mu_i&=&\mu-\epsilon_i\nonumber\\
&\equiv&\mu-\int{\it d}{\bf r}\ V_{\rm t}({\bf r})|w({\bf r}-{\bf r}_i)|^2\simeq\mu-V_{\rm t}({\bf r}_i),
\end{eqnarray}
where $t$ is the hopping parameter for the adjacent sites $i$ and $j$, $U$ is the on-site 
repulsion, and $\mu_i$ is the site dependent chemical potential, given by the sum of the chemical potential $\mu$ and 
the energy offset $\epsilon_i$ of each lattice site due to the slowly varying magnetic trap potential $V_{\rm t}$. 
In Eq.(\ref{bose-hubbard}), the summation in the first term on the right-hand side is restricted to nearest neighbor lattice sites.

\begin{figure}
  \begin{center}
    \scalebox{1.5}[1.5]{\includegraphics{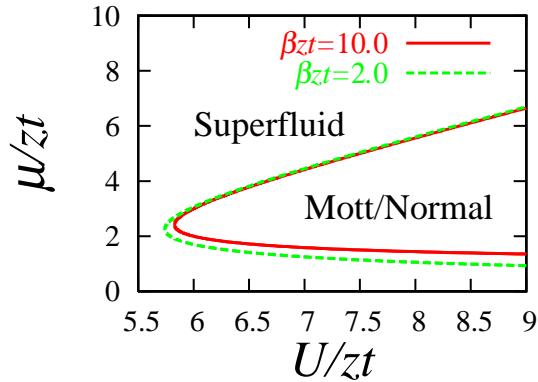}}
    \caption{The phase diagram of the Bose-Hubbard Hamiltonian. The solid line is the phase boundary for $\beta zt=10.0$ and the dashed line is for $\beta zt=2.0$. The Mott phase has the filling factor (average particle number per site) $n=1$.}
    \label{fig:phase}
  \end{center}
\end{figure}

In this paper, we ignore the confining magnetic trap potential $V_{\rm t}$, setting $\mu_i\equiv\mu$ in the Bose-Hubbard model (\ref{bose-hubbard}).
As mentioned in Sec.~\ref{sec:intro}, the properties of the Bose-Hubbard model have been investigated extensively by many authors~\cite{fisher,jaksch,sheshadri,oosten,elstner,kuhner,batrouni,freericks}. 
The effects of the confining trap potential to the equilibrium properties of the Bose-Hubbard model have been studied in Refs.~\cite{batrouni2,batrouni3}.
Here we discuss the superfluid-Mott phase transition in this system using the Landau free energy, which is the function of the order parameter in the superfluid phase defined by $\Psi\equiv\langle b_i\rangle$. 
In Appendix \ref{sec:derivation}, we give a detailed microscopic derivation of the Landau free energy by means of the inversion method developed by Fukuda \textit{et al.}\cite{fukuda}.
The resulting Landau free energy per site is given as 
\begin{eqnarray}\label{mean:landau}
\Gamma(\Psi,\Psi^*)=\Gamma^{(0)}+A|\Psi|^2+B|\Psi|^4,
\end{eqnarray}
where the explicit expressions for $\Gamma^{(0)}$ and the coefficients $A$ and $B$ are given in terms of the parameters $U$, $t$, and $\mu$ in Appendix \ref{sec:derivation}. 
From the free energy in Eq.(\ref{mean:landau}), one finds $\Psi=0$ for $A>0$ and $\Psi\neq 0$ for $A<0$.  Thus the phase boundary between the non-superfluid phase and the superfluid phase with $\Psi\neq 0$ is found by setting $A=0$. 
We note that our Landau theory at $T=0$ reproduces the mean-field results in Refs.~\cite{fisher,sheshadri,oosten}.
Figure \ref{fig:phase} shows the phase diagram for the temperatures $\beta zt=10.0$ and $\beta zt=2.0$. 
The similar phase diagram has been also obtained in Refs.~\cite{dickerscheid,buonsante}
 
Figure \ref{fig:phase} shows that the region of the non-superfluid phase grows as the temperature increases. 
As discussed in Ref.~\cite{dickerscheid}, this does not mean that the region of the Mott insulator phase grows because the Mott insulator phase is meaningful only at zero-temperature. 
On the other hand, one can call the low-temperature non-superfluid phase at $\beta zt=10.0$ a Mott insulator phase for practical purpose , since the compressibility is very close to zero. 
We also note that the phase boundary at $\beta zt=10.0$ is almost identical to that of $T=0$.

In Sec.~\ref{sec:output}, we will investigate excitations of a Bose gas in the optical lattice, modeled by the Bose-Hubbard Hamoltonian in Eq.(\ref{bose-hubbard}), near the phase boundary 
with the fixed value of the chemical potential $\mu/zt\simeq 2.41$ approaching 
from the Mott (or normal) phase by using an output coupling current. This is the main purpose of this paper.

\section{output coupling current expressed by correlation functions}\label{sec:correlation}
In the output coupling experiments, one uses a stimulated Raman transition to change the internal states of atoms from a magnetically trapped state to an untrapped state, and then extracts atoms from a trap to a free space.  
This technique was originally introduced for the atom laser device in order to extract a coherent atomic beam from a Bose-Einstein condensate in a trap potential~\cite{moy}. 
As pointed out in Refs.~\cite{japha,choi,luxat}, this technique can also be used to probe many-body states in a trapped Bose gas. 
In the present paper, we consider the experimental situation of Ref.~\cite{greiner}, where the atoms are initially trapped by using the combined magnetic and optical lattice potentials, as described by $\mathcal{H}_{\rm t}$ in Eq.(\ref{mean:optical}). 
Although we ignore the effect of confining potential in the actual calculations, we still consider the situation where the gas is confined in the finite region due to the magnetic trap potential. 
By using the Raman transition, one can change the internal state of atoms to magnetically untrapped state associated with momentum kick, and push atoms out of the confining region. 
In general, atoms in the output state will feel an optical lattice potential produced by the laser field. 
Following the notation of Ref.~\cite{japha}, for convenience we call the initial and output atomic states ``trapped" and ``free" states respectively, although atoms in both states feel optical lattice potentials.
In this section, we express the output coupling current from a Bose gas in the optical lattice in terms of single-particle correlation functions of the Bose-Hubbard model, following the formalism of Luxat and Griffin \cite{luxat}.

The total Hamiltonian of our system is given by
\begin{eqnarray}\label{hamiltonian}
\mathcal{H}&=&\mathcal{H}_{{\rm t}}+\mathcal{H}_{{\rm f}}+V(t)\\
&=&\mathcal{H}^{0}+V(t), 
\end{eqnarray}
where $\mathcal{H}_{\rm t}$ describes the motion of Bose atoms in a combined magnetic and optical lattice potential and is reduced to the Bose-Hubbard model in Eq.(\ref{bose-hubbard}).
The second term of Eq.(\ref{hamiltonian}) describes the motion of atoms in a free state :
\begin{eqnarray}
\mathcal{H}_{\rm f}&\equiv&\int{\it d}{\bf r}\ \psi_{\rm f}^{\dag}({\bf r})
\left[-\frac{\hbar^2}{2m}\nabla^2+V_{\rm op}'({\bf r})-\mu_{\rm f}\right]\psi_{\rm f}({\bf r})\nonumber\\
&\equiv&\int{\it d}{\bf r}\ \psi_{\rm f}^{\dag}({\bf r})\left[h_{\rm f}({\bf r})-\mu_{\rm f}\right]\psi_{\rm f}({\bf r}),
\end{eqnarray}
where $\psi_{\rm f}({\bf r})$ is the Bose field operator, $\mu_{\rm f}$ is the chemical potential, and 
$V_{{\rm op}}'({\bf r})$ is an optical potential for atoms in this state~\cite{footnote1}.  
We assume that the density of a Bose gas in the free state is so low that the interaction between atoms in the gas can be 
neglected. 
The third term of Eq.(\ref{hamiltonian}) describes the time-dependent probe laser field~\cite{luxat}:
\begin{eqnarray}
V(t)&\equiv&\gamma\int{\it d}{\bf r}
\left[{\it e}^{i({\bf q}\cdot{\bf r}-\omega t)}\psi_{\rm f}^{\dag}({\bf r})\psi_{\rm t}({\bf r})
+{\rm H.c}\right],
\end{eqnarray}
where $\hbar\omega$ is the energy transfer and $\hbar{\bf q}$ is the momentum transfer from the laser field.
We assume that the trapped system and untrapped system do not interact with one another without 
$V(t)$. 
The time dependent external field $V(t)$ couples two systems, acting like the tunneling Hamiltonian.
This field annihilates an atom in one system, while creates an atom in the other system and 
vice versa.
In this process, the laser field causes 
transition between two levels and then one can observe the extracted atoms.
  
We define the output coupling current from a gas in the trapped state to a gas in the free state  
\begin{eqnarray}\label{output_current_definition}
\delta I\equiv-\frac{d\langle\hat{N}_{\rm t}\rangle_{t}}{dt},
\end{eqnarray}
where $\hat{N}_{\rm t}=\int d{\bf r}\psi_{\rm t}^{\dag}({\bf r})\psi_{\rm t}({\bf r})$ and $\langle\cdots\rangle_{t}$ indicates a nonequilibrium statistical average.
The negative sign in this expression implies that the current from a gas in the trapped state to a gas in the free state is positive.
By using the linear response theory regarding the time dependent external field $V(t)$ as a 
perturbation, tunneling current (\ref{output_current_definition}) becomes ~\cite{luxat}
\begin{eqnarray}\label{output_current_general}
\delta I&=&\gamma^2{\rm Re}\int{\it d}{\bf r}\int{\it d}{\bf r}'\ {\it e}^{i{\bf q}\cdot({\bf r}-{\bf r}')}
\int_{-\infty}^{\infty}\frac{{\it d}\omega'}{2\pi}\nonumber\\
& &{}\times\Biggl[
C_{\psi_{\rm t}^{\dag}\psi_{\rm t}}({\bf r}'{\bf r},-\omega')C_{\psi_{\rm f}\psi_{\rm f}^{\dag}}
({\bf r}'{\bf r},\omega'+\omega-\Delta\mu/\hbar)\nonumber\\
& &{}\quad
-C_{\psi_{\rm t}\psi_{\rm t}^{\dag}}({\bf r}{\bf r}',\omega')C_{\psi_{\rm f}^{\dag}\psi_{\rm f}}
({\bf r}{\bf r}',-\omega'-\omega+\Delta\mu/\hbar)\Biggl],
\end{eqnarray}
with the correlation functions defined by
\begin{eqnarray}
C_{\psi^{\dag}\psi}({\bf r}{\bf r}',t)&\equiv&\langle \psi^{\dag}({\bf r}t)\psi({\bf r}',t=0)\rangle\nonumber\\
&\equiv&{\rm Tr}\left\{\rho_{\rm eq}\psi^{\dag}({\bf r}t)\psi({\bf r}')\right\}\nonumber\\
&=&\frac{1}{Z}{\rm Tr}\left\{{\it e}^{-\beta\mathcal{H}^0}
{\it e}^{i\mathcal{H}^0t/\hbar}\psi^{\dag}({\bf r})
{\it e}^{-i\mathcal{H}^0t/\hbar}\psi({\bf r}')\right\}\nonumber\\
& &\left(Z\equiv{\rm Tr}\ {\it e}^{-\beta\mathcal{H}^0}\right),
\end{eqnarray}
where $\rho_{\rm eq}\equiv{\it e}^{-\beta\mathcal{H}^0}$ is 
the initial equilibrium density operator in the absence of the external field, 
$\psi(t)\equiv{\it e}^{-i\mathcal{H}_0t/\hbar}\psi{\it e}^{-i\mathcal{H}_0t/\hbar}$ is the 
field operator in the interaction picture, 
and $\Delta\mu\equiv\mu_{\rm f}-\mu$. 
The Fourier transform of the correlation functions is defined by
\begin{eqnarray}
C_{\psi^{\dag}\psi}({\bf r}{\bf r}',t)=\int_{-\infty}^{\infty}\frac{d\omega}{2\pi}{\it e}^{-i\omega t}C_{\psi^{\dag}\psi}({\bf r}{\bf r}',\omega).
\end{eqnarray}
Using the correlation functions, we define a single-particle spectral function as 
\begin{eqnarray}
A_{\psi^{\dag}\psi}({\bf r}'{\bf r},\omega)
&\equiv&\int_{-\infty}^{\infty}{\it d}t\ {\it e}^{i\omega t}\langle [\psi({\bf r}t),\psi^{\dag}
({\bf r}')]\rangle\nonumber\\
&=&C_{\psi\psi^{\dag}}({\bf r}{\bf r}',\omega)-C_{\psi^{\dag}\psi}({\bf r}'{\bf r},-\omega).
\end{eqnarray}
One can write the correlation functions $C_{\psi^{\dag}\psi}$ and $C_{\psi\psi^{\dag}}$ in terms of this spectral function as~\cite{luxat,kadanoff} 
\begin{eqnarray}
&&C_{\psi^{\dag}\psi}({\bf r}'{\bf r},-\omega)=f(\hbar\omega)A_{\psi^{\dag}\psi}({\bf r}'{\bf r},\omega)\\
&&C_{\psi\psi^{\dag}}({\bf r}{\bf r}',\omega)=\left[1+f(\hbar\omega)\right]A_{\psi^{\dag}\psi}({\bf r}'{\bf r},\omega),
\end{eqnarray}
where $f(\hbar\omega)$ is the Bose distribution function
\begin{eqnarray}
f(\hbar\omega)=\frac{1}{{\it e}^{\beta\hbar\omega}-1}.
\end{eqnarray}

The spectral functions for the non-interacting atoms in an optical lattice can be expressed in terms of the eigenfunction (or Bloch function) $u_{\bf p}({\bf r}){\it e}^{i{\bf p}\cdot{\bf r}}$ and eigenvalue $\epsilon_{\rm f}({\bf p})$ of the single-particle Hamiltonian $h_{\rm f}$~\cite{lifshitz}
\begin{equation}
A_{\psi_{\rm f}^{\dag}\psi_{\rm f}}({\bf r}'{\bf r},\omega)=2\pi
\sum_{{\bf p}}u_{\bf p}({\bf r})u_{\bf p}^*({\bf r}'){\it e}^{-i{\bf p}\cdot({\bf r}'-{\bf r})}
\delta(\omega-[\epsilon_{{\rm f}}({\bf p})-\mu_{\rm f}]/\hbar),\label{spectral_noninteracting}
\end{equation}
where ${\bf p}$ is the quasi-momentum of the Bloch state.
On the other hand, the correlation functions for the trapped atoms are rewritten by using the Wannier functions: 
\begin{eqnarray}
C_{\psi_{\rm t}^{\dag}\psi_{\rm t}}({\bf r}'{\bf r},t)
&=&
\langle \psi_{\rm t}^{\dag}({\bf r}'t)\psi_{\rm t}({\bf r})\rangle\nonumber\\
&=&\sum_{k,l}w^*({\bf r}'-{\bf r}_k)w({\bf r}-{\bf r}_l)
\langle b_k^{\dag}(t)b_l\rangle.
\end{eqnarray}
In order to calculate the correlation functions for the trapped atoms, 
we must calculate the time-correlation functions such as 
$\langle b_k^{\dag}(t)b_l\rangle$. 

For calculating the output-coupling current (\ref{output_current_general}), we need the following four correlation functions:
\begin{eqnarray}
& &C_{\psi_{\rm t}^{\dag}\psi_{\rm t}}({\bf r}'{\bf r},-\omega')\nonumber\\
&&=
\sum_{k,l}w^*({\bf r}'-{\bf r}_k)w({\bf r}-{\bf r}_l)\int_{-\infty}^{\infty}{\it d}t\ 
{\it e}^{-i\omega't}\langle b_k^{\dag}(t)b_l\rangle,\nonumber\\
\label{correlation-1}\\
& &C_{\psi_{\rm t}\psi_{\rm t}^{\dag}}({\bf r}{\bf r}',\omega')\nonumber\\
&&=
\sum_{k,l}w({\bf r}-{\bf r}_k)w^*({\bf r}'-{\bf r}_l)\int_{-\infty}^{\infty}{\it d}t\ 
{\it e}^{i\omega't}\langle b_k(t)b_l^{\dag}\rangle,\nonumber\\
\label{correlation-2}\\
& &C_{\psi_{\rm f}\psi_{\rm f}^{\dag}}({\bf r}'{\bf r},\omega'+\omega-\Delta\mu/\hbar)
\nonumber\\
&&=2\pi\sum_{\bf p}{\it e}^{-i{\bf p}\cdot({\bf r}-{\bf r}')}
\left[1+f(\hbar\omega'+\hbar\omega-\Delta\mu)\right]\nonumber\\
& &{}\times
u_{\bf p}({\bf r}')u_{\bf p}^*({\bf r})
\delta(\omega'+\omega-[\epsilon_{\rm f}({\bf p})-\mu]/\hbar),\nonumber\\
\label{correlation-3}\\
& &C_{\psi_{\rm f}^{\dag}\psi_{\rm f}}({\bf r}{\bf r}',-\omega'-\omega+\Delta\mu/\hbar)
\nonumber\\
&&=2\pi\sum_{\bf p}
{\it e}^{-i{\bf p}\cdot({\bf r}-{\bf r}')}
f(\hbar\omega'+\hbar\omega-\Delta\mu)\nonumber\\
& &{}\times
u_{\bf p}({\bf r}')u_{\bf p}^*({\bf r})
\delta(\omega'+\omega-[\epsilon_{\rm f}({\bf p})-\mu]/\hbar).\nonumber\\
\label{correlation-4}
\end{eqnarray}
The purpose of the next section is to calculate the time-correlation functions 
$\langle b_k^{\dag}(t)b_l\rangle$ and $\langle b_k(t)b_l^{\dag}\rangle$ of the Bose-Hubbard model that appear in the expressions for $C_{\psi_{\rm t}^{\dag}\psi_{\rm t}}$ and $C_{\psi_{\rm t}\psi_{\rm t}^{\dag}}$ given in Eqs.(\ref{correlation-1}) and (\ref{correlation-2}).

\section{correlation functions of the Bose-Hubbard model}
\label{sec:many}
\subsection{Standard-basis operator method for Bose-Hubbard model}
In this section, we apply the standard-basis operator method developed in Ref.~\cite{haley} to calculate the single-particle correlation functions of the Bose-Hubbard model. 
We note that Ref.~\cite{sheshadri} also used this formalism to discuss excitations in the Bose-Hubbard model. 
Here we give detailed calculations of the correlation functions of finite temperatures.
The standard-basis operator is defined as
\begin{equation}
L_{\alpha\alpha'}^{i}\equiv |i\alpha\rangle\langle i\alpha'|,
\end{equation}
where a complete set of functions $\{|i\alpha\rangle\}$ for $i=1,2,...,N_{\rm site}$ and $\alpha=1,2,...,p$ 
consists of the state vectors of atoms at the site $i$ in state $\alpha$. The set of states chosen is arbitrary. Any single-site operator $\mathcal{O}$ 
then can be expressed as
\begin{eqnarray}
\mathcal{O}&=&\sum_i\sum_{\alpha,\alpha'}|i\alpha\rangle\langle i\alpha|\mathcal{O}|i\alpha'\rangle\langle i\alpha'|
\nonumber\\
&=&\sum_i\sum_{\alpha,\alpha'}\langle i\alpha|\mathcal{O}|i\alpha'\rangle L_{\alpha\alpha'}^i.
\end{eqnarray}
Using the standard-basis operator, one can rewrite the Bose-Hubbard Hamiltonian in Eq.(\ref{bose-hubbard}) as
\begin{eqnarray}
\mathcal{H}_{\rm t}&=&-\sum_{<i,j>}\sum_{\alpha,\alpha'}\sum_{\beta,\beta'}T_{\alpha\alpha',\beta\beta'}^{ij}
L_{\alpha\alpha'}^iL_{\beta\beta'}^j+\sum_i\sum_{\alpha,\alpha'}V_{\alpha\alpha'}^iL_{\alpha\alpha'}^i,
\end{eqnarray}
where
\begin{eqnarray}
T_{\alpha\alpha',\beta\beta'}^{ij}&\equiv&
t_{ij}\langle i\alpha|b_i^{\dag}|i\alpha'\rangle\langle j\beta|b_j|j\beta'\rangle\nonumber\\
&&{}
+t_{ij}\langle i\alpha|b_i|i\alpha'\rangle\langle j\beta|b_j^{\dag}|j\beta'\rangle,\label{T}\\
V_{\alpha\alpha'}^{i}&\equiv&\frac{U}{2}\langle i\alpha|\hat{n}^2|i\alpha'\rangle
-\left(\mu_i+\frac{U}{2}\right)\langle i\alpha|\hat{n}|i\alpha'\rangle.\label{V}
\end{eqnarray}

We define the two-time retarded and advanced Green's functions in the coordinate representation in terms of the standard-basis operator by
\begin{eqnarray}\label{green-function}
G_{\alpha\alpha',\beta\beta'}^{R\ ij}(t-t')
&\equiv&\langle\langle L_{\alpha\alpha'}^i(t)|L_{\beta\beta'}^j(t')\rangle\rangle_{R}\nonumber\\
&\equiv&-i\theta(t-t')\langle[L_{\alpha\alpha'}^i(t),L_{\beta\beta'}^j(t')]\rangle,\\
G_{\alpha\alpha',\beta\beta'}^{A\ ij}(t-t')
&\equiv&\langle\langle L_{\alpha\alpha'}^i(t)|L_{\beta\beta'}^j(t')\rangle\rangle_{A}\nonumber\\
&\equiv&i\theta(t'-t)\langle[L_{\alpha\alpha'}^i(t),L_{\beta\beta'}^j(t')]\rangle.
\end{eqnarray}
where $\theta(t)$ is the Heviside step function and $[\ ,\ ]$ denotes the usual commutation relation.
One can interpret the retarded Green's function as the probability amplitude that at time 
$t'$ the $j$th site atom transits from the state $\beta'$ to the state $\beta$ then at time 
$t$ the $i$th atom transits from the state $\alpha'$ to the state $\alpha$.
The Fourier transforms of these Green's functions are defined by
\begin{eqnarray}\label{green-function-omega}
G_{\alpha\alpha',\beta\beta'}^{R(A)\ ij}(\omega)&\equiv&\langle\langle L_{\alpha\alpha'}^i|L_{\beta\beta'}^j\rangle\rangle_{\omega}
\nonumber\\
&\equiv&\int_{-\infty}^{\infty}{\it d}t\ 
{\it e}^{i\omega(t-t')}G_{\alpha\alpha',\beta\beta'}^{R(A)\ ij}(t-t').
\end{eqnarray}
As discussed in Ref.~\cite{zubarev}, one can show that the retarded (advanced) Green's function 
$G^{R}(\omega)$ 
($G^{A}(\omega)$)
can be analytically continued into the region of complex $\omega$ in the upper
(lower) half-plane.
If we make a cut along the real axis, the function
\begin{eqnarray}
G_{\alpha\alpha',\beta\beta'}^{ij}(\omega)
=\left\{
\begin{array}{c}
G_{\alpha\alpha',\beta\beta'}^{R\ ij}(\omega)\quad{\rm for}\quad{\rm Im}\ \omega>0,\\
G_{\alpha\alpha',\beta\beta'}^{A\ ij}(\omega)\quad{\rm for}\quad{\rm Im}\ \omega<0.
\end{array}
\right.
\end{eqnarray}
can be regarded as a single analytic function, consisting of two branches, one of which is defined in the upper and the 
other in the lower half-plane of complex values of $\omega$.

The above Green's functions are related to the time-correlation function through~\cite{zubarev}
\begin{eqnarray}\label{time-correlation}
&&\langle L_{\beta\beta'}^j(t')L_{\alpha\alpha'}^i(t)\rangle\nonumber\\
&&=\int_{-\infty}^{\infty}\frac{{\it d}\omega}{2\pi i}\left\{
G_{\alpha\alpha',\beta\beta'}^{ij}(\omega-i\delta)-G_{\alpha\alpha',\beta\beta'}^{ij}(\omega+i\delta)\right\}\nonumber\\
&&{}\times
f(\hbar\omega){\it e}^{-i\omega(t-t')}\ .
\end{eqnarray}
where $\delta$ is an infinitesimally small positive quantity.
The relation (\ref{time-correlation}) is essential for 
calculating the time-correlation functions in (\ref{correlation-1}) and (\ref{correlation-2}).

Differentiating the Green's function (\ref{green-function-omega}) with respect to $t$ and 
Fourier transforming into $\omega$, we obtain the equation of motion 
\begin{eqnarray}\label{original-eq-of-motion}
\hbar\omega G_{\alpha\alpha',\beta\beta'}^{ij}(\omega)=
\hbar\langle[L_{\alpha\alpha'}^i,L_{\beta\beta'}^j]\rangle
+\langle\langle[L_{\alpha\alpha'}^i,\mathcal{H}_{\rm t}]|L_{\beta\beta'}^j\rangle\rangle_{\omega}.
\end{eqnarray}
This equation is not closed for $G_{\alpha\alpha',\beta\beta'}^{ij}(\omega)$ since it involves the higher-order
Green's function, i.e., 
three-operator Green's function arising from $\langle\langle[L_{\alpha\alpha'}^i,\mathcal{H}_{\rm t}]|L_{\beta\beta'}^j\rangle\rangle_{\omega}$.
Therefore we need to decouple these terms in terms of the two-operator Green's functions. 
For this purpose, we first rewrite Eq.(\ref{original-eq-of-motion}) by using a commutation relation for $L_{\alpha\alpha'}^i$ 
\begin{eqnarray}
[L_{\alpha\alpha'}^i,L_{\beta\beta'}^j]=\delta^{ij}\left(\delta_{\beta\alpha'}L_{\alpha\beta'}^j-\delta_{\alpha\beta'}
L_{\beta\alpha'}^j\right)\ .
\end{eqnarray}
The equation of motion then becomes 
\begin{eqnarray}\label{eq-of-motion}
&&\hbar\omega G_{\alpha\alpha',\beta\beta'}^{ij}(\omega)\nonumber\\
&&=\hbar\langle[L_{\alpha\alpha'}^i,L_{\beta\beta'}^j]\rangle\nonumber\\
&&{}+\sum_{\mu}\left\{V_{\alpha'\mu}^iG_{\alpha\mu,\beta\beta'}^{ij}(\omega)
-V_{\mu\alpha}^iG_{\mu\alpha',\beta\beta'}^{ij}(\omega)\right\}\nonumber\\
& &{}+\sum_l\sum_{\mu}\sum_{\nu\nu'}\Biggl\{T_{\mu\alpha,\nu\nu'}^{il}\langle\langle L_{\mu\alpha'}^iL_{\nu\nu'}^l
|L_{\beta\beta'}^j\rangle\rangle_{\omega}\nonumber\\
&&{}\qquad\qquad\qquad
-T_{\alpha'\mu,\nu\nu'}^{il}\langle\langle L_{\alpha\mu}^iL_{\nu\nu'}^l|L_{\beta\beta'}^j\rangle\rangle_{\omega}\Biggl\}.
\end{eqnarray}

In the random-phase approximation (RPA), the three-operator Green's functions are decoupled by two-operator Green's function as~\cite{haley}
\begin{eqnarray}
&&\langle\langle L_{\mu\alpha'}^iL_{\nu\nu'}^l|L_{\beta\beta'}^j\rangle\rangle_{\omega}\nonumber\\
&&\to
\delta_{\mu\alpha'}\langle L_{\mu\mu}^i\rangle G_{\nu\nu',\beta\beta'}^{lj}(\omega)
+\delta_{\nu\nu'}\langle L_{\nu\nu}^l\rangle G_{\mu\alpha',\beta\beta'}^{ij}(\omega),
\label{rpa-1}\\
&&\langle\langle L_{\alpha\mu}^iL_{\nu\nu'}^l|L_{\beta\beta'}^j\rangle\rangle_{\omega}\nonumber\\
&&\to
\delta_{\alpha\mu}\langle L_{\alpha\alpha}^i\rangle G_{\nu\nu',\beta\beta'}^{lj}(\omega)
+\delta_{\nu\nu'}\langle L_{\nu\nu}^l\rangle G_{\alpha\mu,\beta\beta'}^{ij}(\omega).
\label{rpa-2}
\end{eqnarray}
Substituting Eqs.(\ref{rpa-1}) and (\ref{rpa-2}) into Eq.(\ref{eq-of-motion}), one obtains
\begin{eqnarray}\label{many:inhomogeneous}
&&\hbar\omega G_{\alpha\alpha',\beta\beta'}^{ij}(\omega)\nonumber\\
&&{}=\delta^{ij}\delta_{\alpha\beta'}\delta_{\beta\alpha'}\hbar
(D_{\alpha}^i-D_{\alpha'}^i)\nonumber\\
&&{}
+\sum_l\sum_{\nu\nu'}(D_{\alpha'}^i-D_{\alpha}^i)T_{\alpha'\alpha,\nu\nu'}^{il}G_{\nu\nu',\beta\beta'}^{lj}(\omega)
\nonumber\\
&&{}
+\sum_{\mu}\Biggl[\biggr\{V_{\alpha'\mu}^i-\sum_l\sum_{\nu}D_{\nu}^lT_{\alpha'\mu,\nu\nu}^{il}
\biggr\}
G_{\alpha\mu,\beta\beta'}^{ij}(\omega)\nonumber\\
&&{}\qquad\quad
-\biggr\{V_{\mu\alpha}^i-\sum_l\sum_{\nu}D_{\nu}^lT_{\mu\alpha,\nu\nu}^{il}\biggr\}
G_{\mu\alpha',\beta\beta'}^{ij}(\omega)
\Biggl],
\end{eqnarray}
where $D_{\alpha}^i$ represents the probability that the atom at the site $i$ is in the state $\alpha$, which is defined by
\begin{eqnarray}
D_{\alpha}^i\equiv\langle L_{\alpha\alpha}^i\rangle\ .
\end{eqnarray}

Hereafter we consider a homogeneous system ignoring the trap potential
, i.e., $\mu_i\equiv\mu$. In this case, Eq.(\ref{many:inhomogeneous}) can be transformed into the ${\bf k}$-space:
\begin{eqnarray}\label{eq-of-motion-k}
&&\hbar\omega G_{\alpha\alpha',\beta\beta'}({\bf k},\omega)\nonumber\\
&&{}=\delta_{\alpha\beta'}\delta_{\beta\alpha'}\hbar(D_{\alpha}
-D_{\alpha'})\nonumber\\
&&{}
+\sum_{\nu\nu'}(D_{\alpha'}-D_{\alpha})T_{\alpha'\alpha,\nu\nu'}^k
G_{\nu\nu',\beta\beta'}({\bf k},\omega)\nonumber\\
& &{}+\sum_{\mu}\Biggl[\biggr\{
V_{\alpha'\mu}-\sum_{\nu}D_{\nu}T_{\alpha'\mu,\nu\nu}^0\biggr\}
G_{\alpha\mu,\beta\beta'}({\bf k},\omega)\nonumber\\
&&{}\quad\qquad
-\biggr\{V_{\mu\alpha}-\sum_{\nu}D_{\nu}T_{\mu\alpha,\nu\nu}^0\biggr\}
G_{\mu\alpha',\beta\beta'}({\bf k},\omega)\Biggl],
\end{eqnarray}
where
\begin{eqnarray}
G_{\alpha\alpha',\beta\beta'}^{ij}(\omega)&=&\frac{1}{N_{\rm site}}\sum_{{\bf k}\in {\rm 1stBZ}}
{\it e}^{-i{\bf k}\cdot({\bf r}_i
-{\bf r}_j)}G_{\alpha\alpha',\beta\beta'}({\bf k},\omega),
\end{eqnarray}
with $N_{\rm site}$ being the number of the lattice sites and
\begin{eqnarray}
T_{\alpha\alpha',\beta\beta'}^k&=&\sum_lT_{\alpha\alpha',\beta\beta'}^{lm}
{\it e}^{i{\bf k}\cdot({\bf r}_l-{\bf r}_m)}.
\end{eqnarray}
The summation of the wave vector ${\bf k}$ is restricted to 
the first Brillouin zone.
In order to solve Eq.(\ref{eq-of-motion-k}), we must specify appropriate basis $\{|i\alpha\rangle\}$. 

\subsection{Green's functions}\label{sub:green}
The time-correlation functions in (\ref{correlation-1}) and (\ref{correlation-2}) are expressed in terms of the standard-basis 
operators as
\begin{eqnarray}
\langle b_i^{\dag}(t)b_j\rangle&=&\langle {\it e}^{i\mathcal{H}t/\hbar}b_i^{\dag}{\it e}^{-i\mathcal{H}t/\hbar}b_j\rangle\nonumber\\
&=&\sum_{\alpha\alpha'}\sum_{\beta\beta'}\langle i\alpha|b_i^{\dag}|i\alpha'\rangle\langle j\beta|b_j|j\beta'\rangle
\langle L_{\alpha\alpha'}^i(t)L_{\beta\beta'}^j\rangle,\nonumber\\
\\
\langle b_i(t)b_j^{\dag}\rangle&=&\langle {\it e}^{i\mathcal{H}t/\hbar}b_i
{\it e}^{-i\mathcal{H}t/\hbar}b_j^{\dag}\rangle\nonumber\\
&=&\sum_{\alpha\alpha'}\sum_{\beta\beta'}\langle i\alpha|b_i|i\alpha'\rangle
\langle j\beta|b_j^{\dag}|j\beta'\rangle\langle L_{\alpha\alpha'}^i(t)L_{\beta\beta'}^j\rangle.
\nonumber\\
\end{eqnarray}
Using the relation (\ref{time-correlation}), we find
\begin{eqnarray}
&&\langle b_i^{\dag}(t)b_j\rangle\nonumber\\
&&{}=\sum_{\alpha\alpha'}\sum_{\beta\beta'}\langle i\alpha|b_i^{\dag}|i\alpha'\rangle\langle j\beta|b_j|j\beta'\rangle
\nonumber\\
& &{}\times \int_{-\infty}^{\infty}
\frac{{\it d}\omega}{2\pi i}\ {\it e}^{i\omega t}f(\hbar\omega)\nonumber\\
&&{}\times
\left[G_{\beta\beta',\alpha\alpha'}^{ji}(\omega-i\delta)
-G_{\beta\beta',\alpha\alpha'}^{ji}(\omega+i\delta)\right],\nonumber\\
\label{correlation-real-1}\\
&&\langle b_i(t)b_j^{\dag}\rangle\nonumber\\
&&{}=\sum_{\alpha\alpha'}\sum_{\beta\beta'}\langle i\alpha|b_i|i\alpha'\rangle\langle j\beta|b_j^{\dag}|j\beta'\rangle\nonumber\\
& &{}\times \int_{-\infty}^{\infty}
\frac{{\it d}\omega}{2\pi i}\ {\it e}^{i\omega t}f(\hbar\omega)\nonumber\\
&&{}\times\left[
G_{\beta\beta',\alpha\alpha'}^{ji}(\omega-i\delta)
-G_{\beta\beta',\alpha\alpha'}^{ji}(\omega+i\delta)\right].\nonumber\\
\label{correlation-real-2}
\end{eqnarray}
For solving Eq.(\ref{eq-of-motion-k}) in the Mott insulator phase, we take the Fock states $\{|n\rangle\}$ as the complete set.   
In order to obtain a closed set of equations for the Green's functions, we restrict the Hilbert space to 
$\{|n-1\rangle,|n\rangle, |n+1\rangle\}$, where $n$ is the number of boson per lattice site. 
We note that these states are used in the second-order perturbation calculations in terms of the hopping term at $T=0$~\cite{oosten}. 
With this restricted Hilbert space, Eqs. (\ref{correlation-real-1}) and (\ref{correlation-real-2}) become
\begin{eqnarray}
&&\langle b_i^{\dag}(t)b_j\rangle\nonumber\\
&&{}=\sum_{\alpha\alpha'}\sum_{\beta\beta'}\langle i\alpha|b_i^{\dag}|i\alpha'\rangle\langle j\beta|b_j|j\beta'\rangle
\nonumber\\
& &{}\times \frac{1}{N_{\rm site}}\sum_{{\bf k}\in {\rm 1stBZ}}
\int_{-\infty}^{\infty}\frac{{\it d}\omega}{2\pi i}\ 
{\it e}^{i\{\omega t-{\bf k}\cdot({\bf r}_i-{\bf r}_j)\}}f(\hbar\omega)\nonumber\\
&&{}\times
\left[G_{\beta\beta',\alpha\alpha'}({\bf k},\omega-i\delta)
-G_{\beta\beta',\alpha\alpha'}({\bf k},\omega+i\delta)\right]\nonumber\\
&&{}\equiv\frac{1}{N_{\rm site}}\sum_{{\bf k}\in {\rm 1stBZ}}
\int_{-\infty}^{\infty}\frac{{\it d}\omega}{2\pi i}\ 
{\it e}^{i\{\omega t-{\bf k}\cdot({\bf r}_i-{\bf r}_j)\}}f(\hbar\omega)\nonumber\\
&&{}\times
\left[G_1({\bf k},\omega-i\delta)
-G_1({\bf k},\omega+i\delta)\right],\nonumber\\
\\
&&\langle b_i(t)b_j^{\dag}\rangle\nonumber\\
&&{}=\sum_{\alpha\alpha'}\sum_{\beta\beta'}\langle i\alpha|b_i|i\alpha'\rangle\langle j\beta|b_j^{\dag}|j\beta'\rangle
\nonumber\\
& &{}\times \frac{1}{N_{\rm site}}\sum_{{\bf k}\in {\rm 1stBZ}}
\int_{-\infty}^{\infty}\frac{{\it d}\omega}{2\pi i}\ 
{\it e}^{i\{\omega t-{\bf k}\cdot({\bf r}_i-{\bf r}_j)\}}f(\hbar\omega)\nonumber\\
&&{}\times
\left[G_{\beta\beta',\alpha\alpha'}({\bf k},\omega-i\delta)
-G_{\beta\beta',\alpha\alpha'}({\bf k},\omega+i\delta)\right]\nonumber\\
&&{}\equiv\frac{1}{N_{\rm site}}\sum_{{\bf k}\in {\rm 1stBZ}}
\int_{-\infty}^{\infty}\frac{{\it d}\omega}{2\pi i}\ 
{\it e}^{i\{\omega t-{\bf k}\cdot({\bf r}_i-{\bf r}_j)\}}f(\hbar\omega)\nonumber\\
&&{}\times
\left\{G_2({\bf k},\omega-i\delta)
-G_2({\bf k},\omega+i\delta)\right\},
\end{eqnarray}
where we have introduced the Green's functions $G_1$ and $G_2$ defined in terms of the Bose operators
\begin{eqnarray}
G_1^{ij}(t-t')&\equiv&
\langle\langle b_i^{\dag}(t)|b_j(t')\rangle\rangle,\\
G_2^{ij}(t-t')&\equiv&
\langle\langle b_i(t)|b_j^{\dag}(t')\rangle\rangle,
\end{eqnarray}
and their Fourier transforms
\begin{eqnarray}
G^{ij}_l(\omega)=\frac{1}{N_{\rm site}}\sum_{{\bf k}\in {\rm 1stBZ}}
{\it e}^{-i{\bf k}\cdot({\bf r}_i-{\bf r}_j)}G_l({\bf k},\omega),\quad l=1,2.
\end{eqnarray}
These are related to the Green's functions defined in terms of the standard basis operators 
\begin{eqnarray}
G_1({\bf k},\omega)&\equiv&
\sum_{\alpha\alpha'}\sum_{\beta\beta'}\langle i\alpha|b_i^{\dag}|i\alpha'\rangle\langle j\beta|b_j|j\beta'\rangle
G_{\beta\beta',\alpha\alpha'}({\bf k},\omega),\\
G_2({\bf k},\omega)&\equiv&
\sum_{\alpha\alpha'}\sum_{\beta\beta'}\langle i\alpha|b_i|i\alpha'\rangle\langle j\beta|b_j^{\dag}|j\beta'\rangle
G_{\beta\beta',\alpha\alpha'}({\bf k},\omega).
\end{eqnarray}
Recalling that we restrict our Hilbert space to the Fock space $\{|n-1\rangle,|n\rangle, |n+1\rangle\}$, 
these are expressed as
\begin{eqnarray}
G_1({\bf k},\omega)&=&(n+1)G_{(n)(n+1),(n+1)(n)}({\bf k},\omega)\nonumber\\
&+&
\sqrt{n(n+1)}G_{(n)(n+1),(n)(n-1)}({\bf k},\omega)\nonumber\\
&+&
\sqrt{n(n+1)}G_{(n-1)(n),(n+1)(n)}({\bf k},\omega)\nonumber\\
&+&
nG_{(n-1)(n),(n)(n-1)}({\bf k},\omega),\label{green_1}\\
\nonumber\\
G_2({\bf k},\omega)&=&(n+1)G_{(n+1)(n),(n)(n+1)}({\bf k},\omega)\nonumber\\
&+&\sqrt{n(n+1)}G_{(n)(n-1),(n)(n+1)}({\bf k},\omega)\nonumber\\
&+&\sqrt{n(n+1)}G_{(n+1)(n),(n-1)(n)}({\bf k},\omega)\nonumber\\
&+&nG_{(n)(n-1),(n-1)(n)}({\bf k},\omega),\label{green_2}
\end{eqnarray}
where we omitted the site index of $n$ as we consider a homogeneous system.
Introducing the spectral functions for $G_1$ and $G_2$
\begin{eqnarray}
A_1({\bf k},\omega)&\equiv&-i\left[
G_1({\bf k},\omega-i\delta)-G_1({\bf k},\omega+i\delta)\right],\\
A_2({\bf k},\omega)&\equiv&-i\left[
G_2({\bf k},\omega-i\delta)-G_2({\bf k},\omega+i\delta)\right],
\end{eqnarray}
we obtain the correlation functions as
\begin{eqnarray}
&&\langle b_i^{\dag}(t)b_j\rangle\nonumber\\
&&{}=\frac{1}{N_{\rm site}}\sum_{{\bf k}\in {\rm 1stBZ}}
\int_{-\infty}^{\infty}\frac{{\it d}\omega}{2\pi}\ 
{\it e}^{i\{\omega t-{\bf k}\cdot({\bf r}_i-{\bf r}_j)\}}f(\hbar\omega)A_1({\bf k},\omega),
\nonumber\\
\\
&&\langle b_i(t)b_j^{\dag}\rangle\nonumber\\
&&{}=\frac{1}{N_{\rm site}}\sum_{{\bf k}\in {\rm 1stBZ}}
\int_{-\infty}^{\infty}\frac{{\it d}\omega}{2\pi}\ 
{\it e}^{i\{\omega t-{\bf k}\cdot({\bf r}_i-{\bf r}_j)\}}f(\hbar\omega)A_2({\bf k},\omega).
\nonumber\\
\end{eqnarray}

The explicit forms of the Green's functions (\ref{green_1}) and (\ref{green_2}) are obtained by solving the equation of motion 
(\ref{eq-of-motion-k}). 
The components for $G_1({\bf k},\omega)$ and $G_2({\bf k},\omega)$ in Eqs.(\ref{green_1}) and (\ref{green_2}) are given in Appendix \ref{app:green}.
The poles of these Green's functions determine the dispersion relations of the particle and hole excitations. 
These are given by
\begin{eqnarray}\label{dispersion}
\epsilon_{\rm p,h}({\bf k})
&=&\frac{U}{2}(2n-1)-\mu\nonumber\\
&&{}+\biggr[\frac{(n+1)P_{n+1,n}}{2}
+\frac{nP_{n,n-1}}{2}\biggr]\epsilon({\bf k})\pm\frac{1}{2}\Delta({\bf k}),
\end{eqnarray}
where we denoted $D_{\alpha}-D_{\beta}\equiv P_{\alpha,\beta}$ and 
\begin{eqnarray}\label{gap}
\Delta({\bf k})
&\equiv&
\biggr[U^2+2U\left\{(n+1)P_{n+1,n}-nP_{n,n-1}\right\}\epsilon({\bf k})\nonumber\\
&&{}+\{2n(n+1)P_{n+1,n}P_{n,n-1}+(n+1)^2P_{n+1,n}^2\nonumber\\
&&{}+n^2P_{n,n-1}^2\}\epsilon({\bf k})^2
\biggr]^{1/2}\nonumber\\
&=&\epsilon_{\rm p}({\bf k})-\epsilon_{\rm h}({\bf k}),\\
\epsilon({\bf k})&\equiv& 2t\sum_{j=1}^{3}\cos(k_ja),
\end{eqnarray}
with the lattice constant $a=\lambda/2$. 
\begin{figure}
  \begin{center}
    \begin{tabular}{cc}
      \scalebox{1.0}[1.0]{\includegraphics{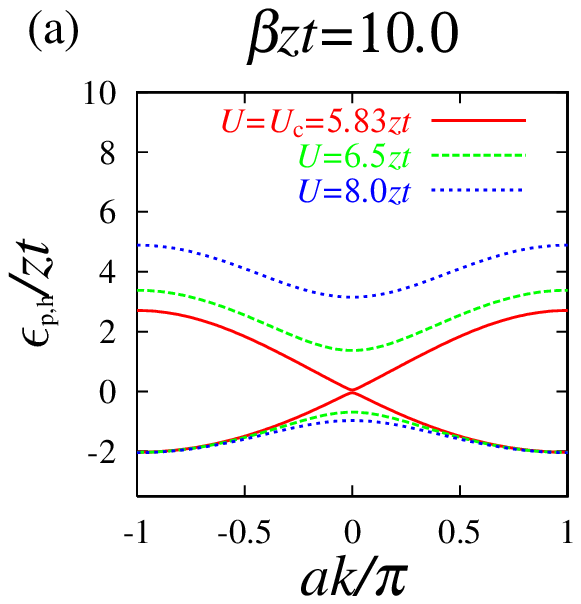}} &
      \scalebox{1.0}[1.0]{\includegraphics{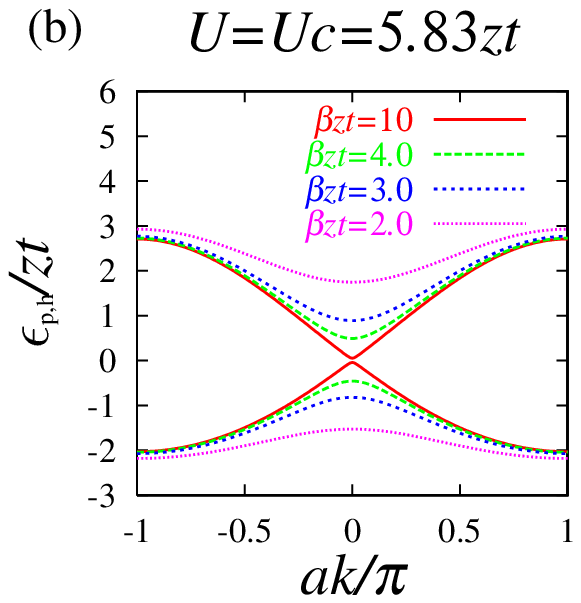}}
    \end{tabular}
    \caption{The dispersion relations of the particle excitation $\epsilon_{\rm p}({\bf k})$ (upper branch) and the hole excitation
    $\epsilon_{\rm h}({\bf k})$ (lower branch) for ${\bf k}=(k,0,0)$. Panel (a) shows the $U$ dependence at the fixed temperature $\beta zt=10.0$, 
    while panel (b) shows the temperature dependence 
    at $U=U_{\rm c}=5.83zt$.}
    \label{fig:dispersion}
  \end{center}
\end{figure}
The dispersion relations in Eq.(\ref{dispersion}) depend on temperature through the function $P$. 
To obtain an explicit expression for the temperature-dependent function $P$, we must determine the occupation probability $D_{\alpha}$ in a self-consistent manner. 
To do this, we follow the procedure described in Sec.~\Roman{six} of Ref.~\cite{haley}. 
This procedure is rather involved, but the final results in the low-temperature limit $D_{n}\gg D_{n\pm1}$ are simply given by (see also Eq.(9.14) of Ref.~\cite{haley})
\begin{eqnarray}
D_n&=&1-\frac{1}{N_{\rm site}}\sum_{{\bf k}\in {\rm 1stBZ}}\left[f(\epsilon_{\rm p}({\bf k}))+f(|\epsilon_{\rm h}({\bf k})|)\right],\\
D_{n+1}&=&\frac{1}{N_{\rm site}}\sum_{{\bf k}\in {\rm 1stBZ}}f(\epsilon_{\rm p}({\bf k})),\\
D_{n-1}&=&\frac{1}{N_{\rm site}}\sum_{{\bf k}\in {\rm 1stBZ}}f(|\epsilon_{\rm h}({\bf k})|),
\end{eqnarray}
where $f(|\epsilon_{\rm h}({\bf k})|)=-1-f(-\epsilon_{\rm h}({\bf k}))$, and
\begin{eqnarray}
P_{n,n+1}&=&1+\frac{1}{N_{\rm site}}\sum_{{\bf k}\in {\rm 1stBZ}}\left[1-2f\left(\epsilon_{\rm p}({\bf k})\right)
+f\left(\epsilon_{\rm h}({\bf k})\right)\right],\label{P_{n,n+1}}\\
P_{n,n-1}&=&1+\frac{1}{N_{\rm site}}\sum_{{\bf k}\in {\rm 1stBZ}}\left[2-f\left(\epsilon_{\rm p}({\bf k})\right)
+2f\left(\epsilon_{\rm h}({\bf k})\right)\right].
\label{P_{n-1,n}}
\end{eqnarray}
The right-hand sides of Eqs.~(\ref{P_{n,n+1}}) and (\ref{P_{n-1,n}}) still depend on $P$ through $\epsilon_{\rm p}({\bf k})$ and $\epsilon_{\rm h}({\bf k})$, and thus we must solve these equations along with (\ref{dispersion}) self-consistently. 
In the very low temperature regime of interest, we can solve these equations by iterating (\ref{P_{n,n+1}}) and (\ref{P_{n-1,n}}) with (\ref{dispersion}).
Hereafter we use the first-order results of this iterative approximation.

Using Eqs.(\ref{app:green_1_1})-(\ref{app:green_1_4}) in Eq.(\ref{green_1}), we obtain the Green's function $G_1$ and the spectral function $A_1$ in the following explicit forms :
\begin{eqnarray}
G_1({\bf k},\omega)
&=&\frac{Z({\bf k})}{\omega-\epsilon_{\rm p}({\bf k})/\hbar}
+\frac{\alpha-Z({\bf k})}
{\omega-\epsilon_{\rm h}({\bf k})/\hbar},\label{green-finite}\\
A_1({\bf k},\omega)&=&2\pi Z({\bf k})\delta\left(\omega-\epsilon_{\rm p}({\bf k})/\hbar\right)
\nonumber\\
&&{}
+2\pi(\alpha-Z({\bf k}))\delta\left(\omega-\epsilon_{h}({\bf k})/\hbar\right),\label{spectral}
\end{eqnarray}
where the wave-function renormalization factor is defined as
\begin{eqnarray}\label{wave-function-renormalization}
Z({\bf k})&=&\frac{1}{\Delta({\bf k})}
\biggr\{(n+1)P_{n,n+1}\left[\mu-U(n-1)\nonumber-nP_{n,n-1}\epsilon({\bf k})\right]\nonumber\\
&&{}
-nP_{n,n-1}\left[\mu-Un+(n+1)P_{n,n+1}\epsilon({\bf k})\right]\nonumber\\
&&{}
+2n(n+1)P_{n,n-1}P_{n,n+1}\epsilon({\bf k})
+\epsilon_{\rm p}({\bf k})\biggr\},
\end{eqnarray}
with 
\begin{eqnarray}
\alpha\equiv(n+1)P_{n,n+1}-nP_{n,n-1}\label{alpha}.
\end{eqnarray}
One can also obtain analogous expressions for $G_2$ and $A_2$.
In order for $A_1$ given in Eq.(\ref{spectral}) to satisfy the sum rule for the spectral function, $\alpha$ must be $1$. 
At $T=0$, one can show that $P_{n,n+1}=P_{n,n-1}=1$ and thus $\alpha=1$. 
However, our numerical evaluation of Eq.(\ref{alpha}) shows that $\alpha$ slightly deviates from $1$ as we increase the temperature. 
This is because we restrict our Hilbert space to $\{|n-1\rangle, |n\rangle, |n+1\rangle\}$. 
In our calculations shown below, we only consider the low temperature region where $\alpha\approx 1$ is satisfied (within 1\%).

At $T=0$, the Green's function and the spectral function reduce to 
\begin{eqnarray}
&&G_1({\bf k},\omega)=\frac{Z({\bf k})}{\omega-\epsilon_{\rm p}({\bf k})/\hbar}
+\frac{1-Z({\bf k})}{\omega-\epsilon_{\rm h}({\bf k})/\hbar},\label{green-T=0}\\
&&A_1({\bf k},\omega)=2\pi Z({\bf k})\delta\left(\omega-\epsilon_{\rm p}({\bf k})/\hbar\right)
\nonumber\\
&&\qquad\qquad\ 
+2\pi(1-Z({\bf k}))\delta\left(\omega-\epsilon_{h}({\bf k})/\hbar\right),
\end{eqnarray}
where the wave-function renormalization factor at $T=0$ is
\begin{eqnarray}
Z({\bf k})=\frac{U(2n+1)-\epsilon({\bf k})+\Delta({\bf k})}{2\Delta({\bf k})},
\end{eqnarray}
and the dispersion relations for the particle and hole excitation at $T=0$ are given by
\begin{eqnarray}
\epsilon_{\rm p,h}(\bf k)&=&\frac{U}{2}(2n-1)-\mu-\frac{\epsilon({\bf k})}{2}\nonumber\\
&&{}
\pm\frac{1}{2}\sqrt{U^2-(4n+2)U\epsilon({\bf k})+\epsilon({\bf k})^2}.
\end{eqnarray}
These results for $T=0$ agree with results in Refs.~\cite{oosten,dickerscheid}.

It is now useful to note the relation between the excitation energies and the phase boundary discussed in Sec.~\ref{sec:bose}. 
The coefficient $A$ in the Landau free energy in Eq.(\ref{mean:landau}) is related to the static susceptibility to the symmetry-breaking external field (see Appendix \ref{sec:derivation}), and thus one has the relation
\begin{eqnarray}\label{landau-green}
A=-\frac{1}{G_1({\bf 0},0)}.
\end{eqnarray}
From the expression for $Z({\bf k})$ in Eq.~(\ref{wave-function-renormalization}), we see that at the transition point $A$ vanishes as $A\propto \Delta_0$, where $\Delta_0\equiv\Delta({\bf k}=0)$ is the energy gap representing the difference between the particle and hole excitations at ${\bf k}=0$. 
Thus, $\Delta_0=0$ signals the phase boundary between the superfluid and Mott (or normal) phase.
In this paper, we have obtained $A$ and $G_1({\bf 0},0)$ independently using the different approaches. 
At $T=0$, it is straightforward to show that Eq.(\ref{landau-green}) holds by comparing Eq.(\ref{green-T=0}) and Eq.~(\ref{A}). 
At finit $T$, we numerically confirmed that the transition point determined from $A=0$ agrees with the vanishing point of $\Delta_0$.

In Fig.\ref{fig:dispersion}, we plot the dispersion relations given in Eq.(\ref{dispersion}). 
With the fixed temperature $\beta zt=10.0$, the energy gap vanishes at $U=U_c\simeq 5.83zt$ and increases as $U/zt$ increases (Fig \ref{fig:dispersion}(a)). 
In Fig.\ref{fig:dispersion}(b) for the fixed $U=U_c$, we find the growth of the energy gap with the increasing temperature. 
These behavior of the energy gap is consistent with the phase diagram in Fig.\ref{fig:phase}. 
As discussed in Sec.\ref{sec:bose}, although there is no Mott insulator phase at finite temperature in a strict sense, we may call the energy gap $\Delta_0$ shown in Fig.\ref{fig:dispersion}(a) for $\beta zt=10.0$ the Mott gap in a practical sense, since the compressibility is very low~\cite{dickerscheid}.

\section{output coupling current}\label{sec:output}
\subsection{General expressions for the output coupling current}
We can now express the output coupling current given in Eq.(\ref{output_current_general}) using the spectral functions derived in the previous section.
The correlation functions (\ref{correlation-1}) and (\ref{correlation-2}) are written as
\begin{eqnarray}
&&C_{\psi_{\rm t}^{\dag}\psi_{\rm t}}({\bf r}'{\bf r},-\omega')\nonumber\\
&&{}=
\frac{1}{N_{\rm site}}\sum_{i,j}w^*({\bf r}'-{\bf r}_i)w({\bf r}-{\bf r}_j)\nonumber\\
&&{}\times
\sum_{{\bf k}\in {\rm 1stBZ}}{\it e}^{i{\bf k}\cdot({\bf r}_i-{\bf r}_j)}f(\hbar\omega')
A_1({\bf k},\omega'),\label{correlation-spectral-1}\\
&&C_{\psi_{\rm t}\psi_{\rm t}^{\dag}}({\bf r}{\bf r}',\omega')\nonumber\\
&&{}=
\frac{1}{N_{\rm site}}\sum_{i,j}w({\bf r}-{\bf r}_i)w^*({\bf r}'-{\bf r}_j)\nonumber\\
&&\times
\sum_{{\bf k}\in {\rm 1stBZ}}{\it e}^{i{\bf k}\cdot({\bf r}_i-{\bf r}_j)}f(-\hbar\omega')
A_2({\bf k},-\omega').\label{correlation-spectral-2}
\end{eqnarray}
Using Eqs.(\ref{correlation-spectral-1}) and (\ref{correlation-spectral-2}) along with Eqs.(\ref{correlation-3}) and (\ref{correlation-4}) in Eq.(\ref{output_current_general}) and introducing the function $\phi({\bf q},{\bf p})$ as 
\begin{eqnarray}
\phi({\bf q},{\bf p})\equiv\int{\it d}{\bf r}~w({\bf r})u_{\bf p}^*({\bf r}){\it e}^{i({\bf q}-{\bf p})\cdot{\bf r}},
\end{eqnarray}
we obtain the following expression for the output-coupling current:
\begin{widetext}
\begin{eqnarray}\label{momentum-resolved-wannier}
\delta I&=&\frac{(2\pi)^6\gamma^2}{N_{\rm site}^2a^9}
\sum_{{\bf k}\in {\rm 1stBZ}}\sum_{\bf p}
|\phi({\bf q},{\bf p})|^2
\nonumber\\
& &{}\times\Biggl\{\delta\left({\bf k}+{\bf p}-{\bf q}\right)
\left[1+f(\epsilon_{\rm f}({\bf p})-\mu_{\rm f})\right]
f(\epsilon_{\rm f}({\bf p})-[\hbar\omega+\mu])
A_1({\bf k},\omega-[\epsilon_{{\bf p}{\rm n}}-\mu]/\hbar)\nonumber\\
& &{}\quad -\delta\left({\bf p}-{\bf k}-{\bf q}\right) 
f(\epsilon_{\rm f}({\bf p})-\mu_{\rm f})
\left[1+f(\epsilon_{\rm f}({\bf p})-[\hbar\omega+\mu])\right]
A_2({\bf k},-\omega+[\epsilon_{\rm f}({\bf p})-\mu]/\hbar)\Biggl\}.\nonumber\\
\end{eqnarray}
\end{widetext}
In Eq.(\ref{momentum-resolved-wannier}), the $\delta$-functions represents the conservation of the total momentum. 
The first term in Eq.(\ref{momentum-resolved-wannier}) describes a current of atoms going out from the combined trap, while the second term describes a current of atoms going into the combined trap.

When we assume that $f(\epsilon_{\rm f}({\bf p})-\mu_{\rm f})\ll 1$, which results in negligible tunneling from untrapped atomic gas back to the combined trap region, the above expression reduces to
\begin{widetext}
\begin{eqnarray}\label{current}
\delta I&=&\frac{(2\pi)^6\gamma^2}{N_{\rm site}^2a^9}
\sum_{{\bf k}\in {\rm 1stBZ}}\sum_{\bf p}
|\phi({\bf q},{\bf p})|^2
\nonumber\\
& &{}\times\delta\left({\bf k}+{\bf p}-{\bf q}\right)
f(\epsilon_{\rm f}({\bf p})-[\hbar\omega+\mu])
A_1({\bf k},\omega-[\epsilon_{\rm f}({\bf p})-\mu]/\hbar)
\end{eqnarray}
\end{widetext}
We have thus arrived at the general form of the output-coupling current from an optical lattice in the non-superfluid (i.e., Mott insulator or normal) phase in terms of the spectral function.

\subsection{Numerical evaluation}
We now calculate the output current given by Eq.~(\ref{current}) numerically.
The sums over ${\bf p}$ and ${\bf k}$ in Eq.~(\ref{current}) can be now replaced with the integrals through 
$\sum_{{\bf p}}\rightarrow V\int{\it d}{\bf p}/{(2\pi)^3}$ and 
$\sum_{{\bf k}\in {\rm 1stBZ}}\rightarrow V\int_{-\pi/a}^{\pi/a}{\it d}{\bf k}/{(2\pi)^3}$,
where $V=N_{\rm site}a^3$ is the volume of the system and $a$ is the lattice constant.  
Then the output coupling current becomes
\begin{widetext}
\begin{eqnarray}
\delta I
&=&\frac{(2\pi)^4\gamma^2}{a^3}
\int\frac{{\it d}{\bf p}}{(2\pi)^3}
\int_{-\pi/a}^{\pi/a}{\it d}{\bf k}|\phi({\bf q},{\bf p})|^2
\delta({\bf k}+{\bf p}-{\bf q})\nonumber\\
&&{}\times
\biggr[
Z({\bf k})
\delta(\epsilon_{\rm f}({\bf p})/\hbar-\epsilon_{\rm p}({\bf k})/\hbar-[\omega+\mu/\hbar])
f(\epsilon_{\rm p}({\bf k}))\nonumber\\
&&{}+(\alpha-Z({\bf k}))
\delta(\epsilon_{\rm f}({\bf p})/\hbar-\epsilon_{\rm h}({\bf k})/\hbar-[\omega+\mu/\hbar])
f(\epsilon_{\rm h}({\bf k}))
\biggr],\label{current-final-1}
\end{eqnarray}
\end{widetext}
where  
$\epsilon_{\rm p,h}({\bf k})$ and $\alpha$ have been given in Eqs.(\ref{dispersion}) and 
(\ref{alpha}), respectively.

Let us now define the momentum-resolved current, which describes each ${\bf p}$ process, by
\begin{widetext}
\begin{eqnarray}\label{momentum-resolved}
\delta I({\bf p})
&\equiv&\frac{(2\pi)^4\gamma^2}{a^3}
\int_{-\pi/a}^{\pi/a}{\it d}{\bf k}|\phi({\bf q},{\bf p})|^2
\delta({\bf k}+{\bf p}-{\bf q})\nonumber\\
&&{}\times
\biggr[
Z({\bf k})
\delta(\epsilon_{\rm f}({\bf p})/\hbar-\epsilon_{\rm p}({\bf k})/\hbar-[\omega+\mu/\hbar])
f(\epsilon_{\rm p}({\bf k}))\nonumber\\
&&{}+(\alpha-Z({\bf k}))
\delta(\epsilon_{\rm f}({\bf p})/\hbar-\epsilon_{\rm h}({\bf k})/\hbar-[\omega+\mu/\hbar])
f(\epsilon_{\rm h}({\bf k}))
\biggr].
\end{eqnarray}
\end{widetext} 
We will show that the momentum-resolved current defined in Eq.~(\ref{momentum-resolved}) exhibits a very interesting feature, while we may not be able to see any remarkable character by observing the total output coupling current in Eq.(\ref{current-final-1}).
The physical meaning of 
Eq.(\ref{momentum-resolved}) is the weighted current with conserving the 
total momentum and the total energy.
It describes two energy-conserving processes.  
In the first process, the energies of a photon from the external laser field and a particle excitation is 
transferred to a free atom
\begin{eqnarray}
\hbar\omega+\mu+\epsilon_{\rm p}({\bf k})=\epsilon_{\rm f}({\bf p}).
\end{eqnarray}
In the second process, the energies of photon from the external laser field and a hole excitation is transferred to a free atom
\begin{eqnarray}
\hbar\omega+\mu+\epsilon_{\rm h}({\bf k})=\epsilon_{\rm f}({\bf p}).
\end{eqnarray}
These processes are weighted by the wave function renormalization and the Bose distribution 
function. 
Since we are interested in probing the low-energy excitations in the optical lattice, we evaluate the momentum-resolved current for the small momentum ${\bf p}=(0,0.01,0.01)\pi/a$ and ${\bf q}={\bf 0}$ as a function of the energy transfer  $\hbar\omega$ from the laser field. 
For simplicity, we approximate the low-energy single-particle eigenstate of the free gas by the plane wave, i.e., 
\begin{eqnarray}
u_{\bf p}({\bf r})=\frac{1}{\sqrt{V}},\qquad \epsilon_{\rm f}({\bf p})=\frac{\hbar^2{\bf p}^2}{2m}.
\end{eqnarray}

\begin{figure}
  \begin{center}
    \begin{tabular}{cc}
      \scalebox{1.0}[1.0]{\includegraphics{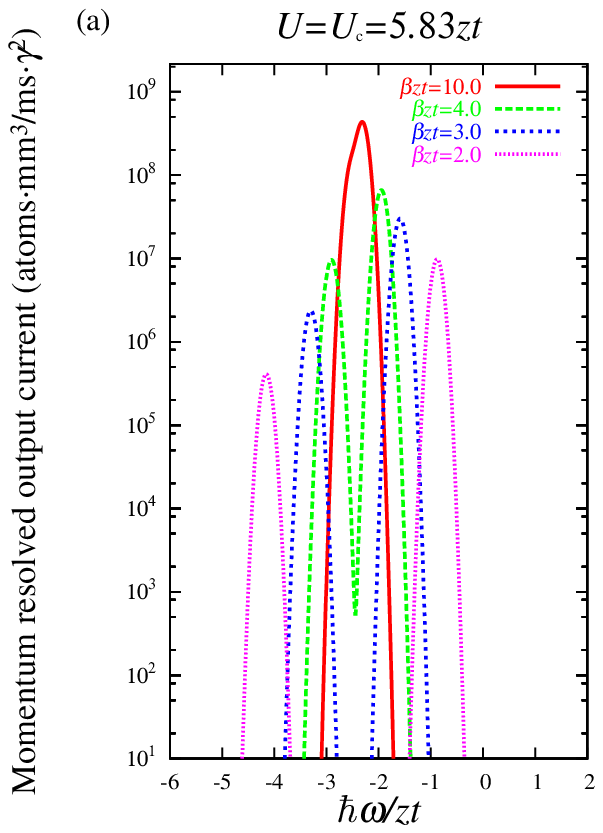}} &
      \scalebox{1.0}[1.0]{\includegraphics{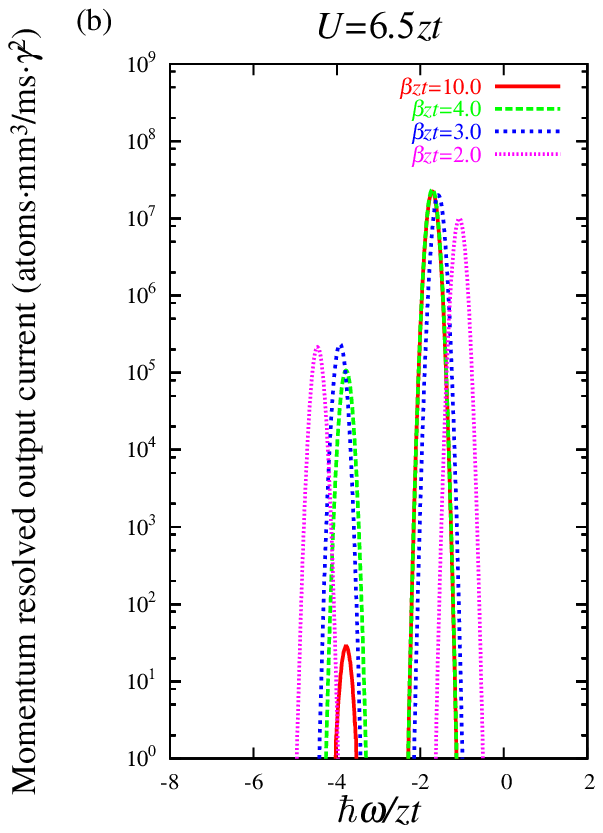}}\\
      \scalebox{1.0}[1.0]{\includegraphics{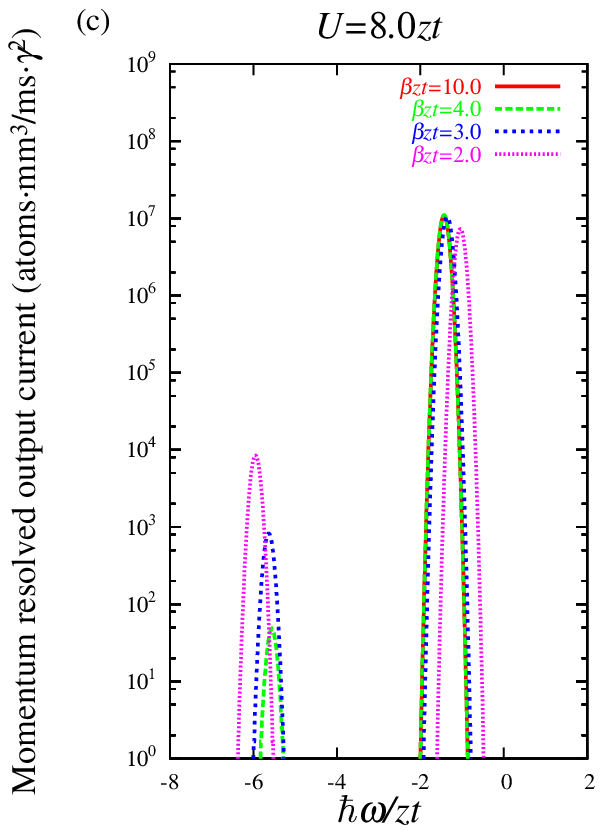}} &
      \scalebox{1.0}[1.0]{\includegraphics{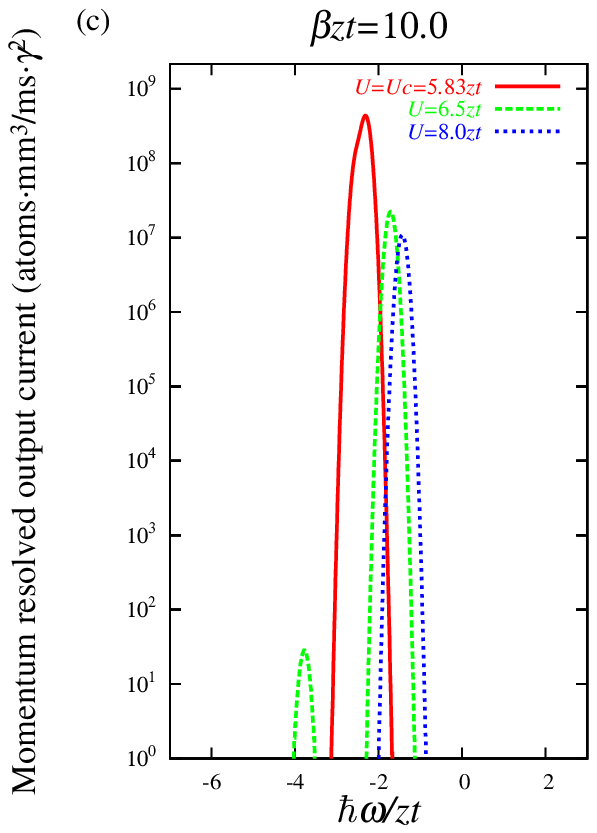}}
    \end{tabular}
    \caption{Momentum resolved output coupling current for ${\bf p}=(0,0.01,0.01)\pi/a$ and ${\bf q}=0$. Panels (a)-(c) are plots with the fixed values of $U$ and various temperatures $\beta zt=10.0,\ 4.0,\ 3.0,\ 2.0$. Panel (d) is a plot with the fixed temperature with $U/zt=5.83,\ 6.5,\ 8.0$. Here we take $\lambda=852$nm and $m$ and $a_s$ for $^{87}$Rb atom, and $V_0$ to adjust the ratio $U/zt$}
    \label{fig:current-U}
  \end{center}
\end{figure}

In Fig.\ref{fig:current-U}, we plot the momentum-resolved current for various temperatures with fixed values of $U$ (Figs.\ref{fig:current-U}(a)-(c)), and for various values of $U$ with a fixed temperature (Fig.\ref{fig:current-U}(d)). 
Note the logarithmic scale for the atom current.
In calculating Eq.(\ref{momentum-resolved}), we approximate the Wannier function $w$ by the Gaussian wave function~\cite{oosten,tsuchiya}.
When $U>U_{\rm c}$, the energy spectrum of the current has two peaks at any temperature. 
Obviously, the lower peak is contributed from the particle-excitation $\epsilon_{\rm p}({\bf k})$ 
and the upper peak is contributed from the hole excitation $\epsilon_{\rm h}({\bf k})$.
Thus, the distance between the two peaks provides the energy gap $\Delta_0$.

From Figs.\ref{fig:current-U}(a)-(c), we find two characteristic features in the temperature dependence of the output coupling current. First, the distance between the two peaks increases as the temperature increases, reflecting the growth of the energy gap in the excitation spectra as shown in Fig.\ref{fig:dispersion}(a). 
Second, the peak corresponding to the particle excitation grows as the temperature increases, reflecting the thermally populated particle excitations. 
One can understand the temperature dependence of the particle and hole peaks from the expression for the momentum-resolved current in Eq.(\ref{momentum-resolved}). 
The particle-excitation peak exists only at finite temperatures since $f(\epsilon_{\rm p}({\bf k}))\to 0$ as $T\to0$. 
In contrast, the hole-excitation peak exists both at zero and finite temperatures since $f(\epsilon_{\rm h}({\bf k}))=-1-f(|\epsilon_{\rm h}({\bf k})|)\to -1$ as $T\to 0$. 
This means that one can only detect the particle excitations and hence the energy gap at finite temperatures. 

In Fig.\ref{fig:current-U}(d) with the fixed temperature $\beta zt=10.0$, we find that the particle and hole peaks merge into a single peak at $U=U_c$, where the Mott gap vanishes. 
As shown in Fig.\ref{fig:current-U}(a), with fixing $U=U_c$ and increasing the temperature, the peak is split into two peaks. 
This reflects the existence of the energy gap at finite temperatures even at $U=U_c$ as shown in Fig.\ref{fig:dispersion}(b). 
This gap, however, is not associated with the Mott insulator phase, as one is in the normal phase away from the critical point. 

We now comment on the experimental observability of the particle-hole gap in the output coupling current. 
In most cases the contribution in the current from the particle excitations is much smaller than the contribution from the hole excitations. 
However, the two peaks corresponding the particle and hole excitations are very sharp and do not overlap with each other except very close to the critical point. 
Therefore, the particle peaks should be visible as a distinct contribution, although high-precision experiments may be required.

\section{conclusions}\label{sec:conclusions}
In this paper, we have studied the single-particle excitations in a Bose gas in an optical lattice. 
We calculated the single-particle Green's functions of the Bose-Hubbard model in the Mott insulator phase 
by using the standard basis operator formalism whose poles provide 
the dispersion relations of particle and hole excitations. 
These dispersion relations show that the energy gap between the particle and hole excitations at ${\bf k}=0$ vanishes at the critical point. 
The energy gap increases with the increasing temperature or increasing $U/zt$. 
The same behavior of the energy gap has also been found in Refs.~\cite{oosten,dickerscheid}. 

In order to gain information of the single-particle excitations in the Mott phase, 
we have calculated the output coupling current using the linear-response formalism given by Luxat and Griffin~\cite{luxat}.
We have shown that the momentum-resolved current has two distinct contributions from the particle and hole excitations, and thus one can directly measure the Mott gap.  
Moreover, this can be used to detect the transition point from the Mott insulator phase to the 
superfluid phase, where the Mott gap vanishes. 
We note that the contribution from the particle excitations is strongly temperature dependent, vanishing at $T=0$.
In most cases the particle-excitation contribution is much smaller than the hole-excitation contribution. 
Nevertheless we expect the particle-excitation peak to be experimentally observable since it is distinct from the hole peak, as shown in Fig.\ref{fig:current-U}.

In the present work we have restricted our calculations to the Mott insulator phase. 
It is next important to study how the output current changes as we enter the superfluid phase. 
In the superfluid phase, we expect to have a direct tunneling current from the condensate component, as well as processes involving quasi-particle excitations~\cite{japha,choi,luxat}. 
Extending the present study to the superfluid phase will be given in a future work.

Finally, we mention the recent experiment by St\"{o}ferle \textit{et al}.\cite{stoferle} investigating excitations in a trapped Bose gas in an optical lattice using the Bragg spectroscopy. 
This type of experiment deals with the density-fluctuation spectrum~\cite{oosten2,ana,greiner}. 
In contrast, our present paper probes the single-particle Green's functions in a Bose gas in an optical lattice, which is more directly related to the single particle and hole excitations that contain important information about the superfluid-Mott insulator transition.

\begin{acknowledgements}
We acknowledge David Luxat and Chikara Ishii for useful discussions.
\end{acknowledgements}

\appendix
\section{Derivation of the Landau free energy by using the inversion method}\label{sec:derivation}
This Appendix gives a detailed derivation of the Landau free energy for the Bose-Hubbard model used in Sec.~\ref{sec:bose}. 
Our aim is to derive a free energy as a function of an order parameter describing the superfluid phase, defined by $\Psi_i\equiv\langle b_i\rangle$, by perturbative expansion in the hopping term. 
Following Ref.~\cite{fukuda} (see also Refs.~\cite{georges,shiba}), we introduce an expansion parameter $g$ and write the Hamiltonian as
\begin{eqnarray}\label{b:H_0+H_1}
\mathcal{H}(g)\equiv\mathcal{H}_0(g)+g\mathcal{H}_1,
\end{eqnarray}
where
\begin{eqnarray}
\mathcal{H}_0(g)&=&\frac{U}{2}\sum_ib_i^{\dag}b_i^{\dag}b_ib_i-\sum_i\mu_ib_i^{\dag}b_i
\nonumber\\
&&{}-\sum_i[\eta_i^*(g)b_i+\eta_i(g)b_i^{\dag}],\label{b:H_0}\\
\mathcal{H}_1&=&-t\sum_{<i,j>}b_i^{\dag}b_j\label{b:H_1}.
\end{eqnarray}
In Eq.(\ref{b:H_0}), we have introduced the symmetry-breaking external fields $\eta$ and $\eta^*$, which controls the order parameter $\Psi_i$. 
We regard $\eta$ and $\eta^*$ as functions of the coupling constant $g$, imposing the condition that the relation 
\begin{eqnarray}\label{d:psi-definition}
\Psi_i=\frac{{\rm Tr}\left[{\it e}^{-\beta\mathcal{H}(g)}b_i\right]}{{\rm Tr}{\it e}^{-\beta\mathcal{H}(g)}}
\end{eqnarray}
should hold independent of $g$. 

The thermodynamic potential associated with the Hamiltonian Eq.(\ref{b:H_0+H_1}) is given by
\begin{equation}\label{d:omega}
\Omega[\eta,\eta^*]=-\frac{1}{\beta}\ln\mbox{Tr}\ {\it e}^{-\beta \mathcal{H}(g)}.
\end{equation}
By performing the Legendre transformation, the Helmholtz free energy as a function of the order parameters $\Psi$ and $\Psi^*$ is given as
\begin{equation}\label{d:legendre}
\Gamma[\Psi,\Psi^*]=\Omega[\eta,\eta^*]
+\sum_i[\eta_i^*(g)\Psi_i+\eta_i(g)\Psi^*_i].
\end{equation}
Differentiating the above equation in regard to $\Psi^*$, we find
\begin{eqnarray}
\frac{\partial\Gamma[\Psi,\Psi^*]}{\partial\Psi_i^*}=\eta_i[g;\Psi_i,\Psi_i^*].\label{d:gamma}
\end{eqnarray}
The required free energy is obtained by setting $g=1$ and turning off the artificial external fields, i.e., $\eta_i(g=1)=0$ at the end of the calculation. 
Then, the equation that determines $\Psi_i$ is given by
\begin{equation}
\frac{\partial\Gamma[\Psi,\Psi^*]}{\partial\Psi_i^*}=0.\label{d:eq-of-phi}
\end{equation}

We now expand the thermodynamic potential $\Omega$ in powers of $g$ by using 
\begin{eqnarray}
{\it e}^{-\beta\Omega[g;\eta,\eta^*]}&=&\mbox{Tr}\ {\it e}^{-\beta\mathcal{H}(g)}\nonumber\\
&=&\sum_{n=0}^{\infty}\left(-\frac{g}{\hbar}\right)^n\frac{1}{n!}
\int_0^{\beta\hbar}d\tau_1\cdots\int_0^{\beta\hbar}d\tau_n
\mbox{Tr}\ \left\{{\it e}^{-\beta H_0(0)}\mbox{T}_{\tau}\ [\mathcal{H}_1(\tau_1)\cdots \mathcal{H}_1(\tau_n)]\right\}\nonumber\\
&=&{\rm Tr}\ {\it e}^{-\beta\mathcal{H}_0(0)}
+\left(-\frac{g}{\hbar}\right)\int_0^{\beta\hbar}d\tau_1\ {\rm Tr}\left[{\it e}^{-\beta\mathcal{H}_0(0)}\mathcal{H}(\tau_1)\right]\nonumber\\
& &{}+\left(-\frac{g}{\hbar}\right)^2\int_0^{\beta\hbar}d\tau_1\int_0^{\tau_1}d\tau_2{\rm Tr}\left[{\it e}^{-\beta\mathcal{H}_0(0)}\mathcal{H}(\tau_1)\mathcal{H}_1(\tau_2)\right]+\cdots,
\end{eqnarray}
where $\mathcal{H}_1(\tau)\equiv {\it e}^{\mathcal{H}_0(0)\tau/\hbar}\mathcal{H}_1{\it e}^{-\mathcal{H}_0(0)\tau/\hbar}$, and the operator ${\rm T}_{\tau}$ is the imaginary time ordering operator. 
We expand $\Omega[g;\eta,\eta^*]$ as
\begin{eqnarray}
\Omega[g;\eta,\eta^*]
=\Omega^{(0)}[\eta,\eta^*]+g\Omega^{(1)}[\eta,\eta^*]
+g^2\Omega^{(2)}[\eta,\eta^*]+\cdots.
\end{eqnarray}
In our formalism, $\eta$ is also expanded through the relation (\ref{d:psi-definition}) as
\begin{eqnarray}
\eta[g;\Psi,\Psi^*]=\eta^{(0)}[\Psi,\Psi^*]+g\eta^{(1)}[\Psi,\Psi^*]+g^2\eta^{(2)}[\Psi,\Psi^*]+\cdots.
\end{eqnarray} 
By the Legendre transformation, the Helmholtz free energy is expanded as
\begin{eqnarray}
\Gamma[\Psi,\Psi^*]=\Gamma^{(0)}[\Psi,\Psi^*]+g\Gamma^{(1)}[\Psi,\Psi^*]+g^2\Gamma^{(2)}[\Psi,\Psi^*]+\cdots.
\end{eqnarray}

The zeroth-order term $\Gamma^{(0)}$ is given by
\begin{eqnarray}
\Gamma^{(0)}[\Psi,\Psi^*]=\Omega^{(0)}[\eta^{(0)},\eta^{*(0)}]+\sum_i\left[\eta_i^{(0)}\Psi_i^*+\eta_i^{*(0)}\Psi_i\right]\label{d:gamma-0th},
\end{eqnarray}
where $\eta^{(0)}$ is related to $\Psi$ through (\ref{d:psi-definition}) with $g=0$. 
The next two terms are given by
\begin{eqnarray}
\Gamma^{(1)}[\Psi,\Psi^*]&=&\Omega^{(0)}[\eta^{(0)},\eta^{*(0)}]
+\sum_i\left\{\frac{\partial\Omega^{(0)}[\eta^{(0)},\eta^{*(0)}]}{\partial\eta_i^{(0)}}\eta_i^{(1)}
+\frac{\partial\Omega^{(0)}[\eta^{(0)},\eta^{*(0)}]}{\partial\eta_i^{*(0)}}\eta_i^{*(1)}\right\}\nonumber\\
&&{}+\sum_i\left[\eta_i^{(1)}\Psi_i^*+\eta_i^{*(1)}\Psi_i\right]\nonumber\\
&=&\Omega^{(1)}[\eta^{(0)},\eta^{*(0)}]\label{d:gamma-1st}\\
\Gamma^{(2)}[\Psi,\Psi^*]&=&\Omega^{(2)}[\eta^{(0)},\eta^{*(0)}]\nonumber\\
&&{}+\frac{1}{2}\sum_i\biggr\{
\frac{\partial\Omega^{(1)}[\eta^{(0)},\eta^{*(0)}]}{\partial\eta_i}\eta_i^{(1)}
+\frac{\partial\Omega^{(1)}[\eta^{(0)},\eta^{*(0)}]}{\partial\eta_i^*}\eta_i^{*(1)}\biggr\}.
\label{d:gamma-2nd}
\end{eqnarray}
Following the procedure described in Ref.~\cite{fukuda}, one can successively calculate $\eta^{(n)}$, and hence systematically expand $\Gamma$ in powers of the coupling constant $g$.

We now explicitly derive the free energy as a function of the order parameter to first order in the coupling constant $g$. 
The zeroth-order term of the thermodynamic potential is given by
\begin{eqnarray}
\Omega^{(0)}[\eta^{(0)},\eta^{*(0)}]&=&-\frac{1}{\beta}\ln\mbox{Tr}\ {\it e}^{-\beta \mathcal{H}(0)}\nonumber\\
&=&-\frac{1}{\beta}\ln\mbox{Tr}\ {\it e}^{-\beta \sum_i\mathcal{H}_i}\nonumber\\
&\equiv&\sum_i\Omega_i[\eta^{(0)},\eta^{*(0)}],
\end{eqnarray}
where we have defined the local Hamiltonian as
\begin{eqnarray}
\mathcal{H}_i\equiv\frac{U}{2}b_i^{\dag}b_i^{\dag}b_ib_i-\mu_ib_i^{\dag}b_i
-\left(\eta_i^{*(0)}b_i+\eta_i^{(0)}b_i^{\dag}\right).
\end{eqnarray}
As noted above, $\eta_i^{(0)}$ is a function of $\Psi_i$ through the relation (\ref{d:psi-definition}), but it is in general difficult to obtain an explicit functional form of $\eta_i^{(0)}[\Psi_i,\Psi_i^*]$. 
In the spirit of the Landau free energy, we assume that the order parameter is small and thus we expand $\Omega_i$ in powers of $\eta_i^{(0)}$ and $\eta_i^{*(0)}$ as follows:
\begin{eqnarray}
\Omega_i^{(0)}[\eta^{(0)},\eta^{*(0)}]
&\simeq&-\frac{1}{\beta}\ln Z_0
-\langle f_i\rangle|\eta_i^{(0)}|^2\nonumber\\
&&{}
+\left\{\langle h_i\rangle
+\frac{\beta}{2}\left[\langle f_i^2\rangle-\langle f_i\rangle^2\right]\right\}
|\eta_i^{(0)}|^4,
\label{d:omega_0}
\end{eqnarray}
where we have introduced $f$ and $h$ as
\begin{eqnarray}
f&=&\frac{n+1}{Un-\mu_i}-\frac{n}{U(n-1)-\mu_i},\label{f}\\
h&=&
\frac{(n+1)(n+2)}{(Un-\mu_i)^2[U(2n+1)-2\mu_i]}
-\frac{(n+1)^2}{(Un-\mu_i)^3}\nonumber\\
&&{}-\frac{n^2}{[U(n-1)-\mu_i]^3}
-\frac{n(n+1)}{(Un-\mu_i)^2[U(n-1)-\mu_i]}\nonumber\\
& &{}-\frac{n(n-1)}{[U(n-1)-\mu_i]^2[U(2n-3)-2\mu_i]}\nonumber\\
&&{}
-\frac{n(n+1)}{(Un-\mu_i)[U(n-1)-\mu_i]^2},\label{h}
\end{eqnarray}
and defined the local thermal average by
\begin{eqnarray}
& &Z_0\equiv \sum_n{\it e}^{-\beta\epsilon_n},\label{thermal-average}\\
& &\langle\cdots\rangle\equiv \frac{1}{Z_0}\sum_n(\cdots){\it e}^{-\beta\epsilon_n},
\end{eqnarray}
with
\begin{eqnarray}
\epsilon_n\equiv\frac{U}{2}n(n-1)-\mu_in.
\end{eqnarray}
Differentiating $\Omega^{(0)}[\eta^{(0)},\eta^{*(0)}]$ with respect to $\eta_i^*$, we obtain 
$\Psi_i=\Psi_i^{(0)}[\eta^{(0)},\eta^{*(0)}]$
\begin{eqnarray}\label{d:phi}
\Psi_i&=&-\frac{\partial \Omega^{(0)}[\eta^{(0)},\eta^{*(0)}]}{\partial\eta_i^*}\nonumber\\
&=&\langle f\rangle\eta_i^{(0)}
+\left\{2\langle g(n)\rangle
+\beta\left[\langle f^2\rangle-\langle f\rangle^2\right]\right\}
\eta_i^{(0)}|\eta_i^{(0)}|^2\nonumber\\
&\equiv&\Psi_i^{(0)}[\eta^{(0)},\eta^{*(0)}].
\end{eqnarray}
The external fields are expressed by the order parameters by inverting Eq. (\ref{d:phi}) as 
\begin{eqnarray}\label{app:eta^{(0)}}
\eta_i^{(0)}[\Psi]=(\Psi^{(0)})^{-1}[\Psi].
\end{eqnarray}

Next, we calculate the 1st order term of the Landau free energy. From Eq. (\ref{d:gamma-1st}), we find
\begin{eqnarray}
\Gamma^{(1)}[\Psi,\Psi^*]=\langle\mathcal{H}_1\rangle_{g=0}=-t\sum_{<i,j>}\langle b_i^{\dag}b_j\rangle_{g=0},
\end{eqnarray}
where the expectation $\langle\ \rangle_{g=0}$ is evaluated with respect to $\mathcal{H}(0)$. 
We find that
\begin{equation}\label{d:gamma-1st-final}
\Gamma^{(1)}[\Psi, \Psi^*]=-t\sum_{<i,j>}\Psi_i^*\Psi_j.
\end{equation}
Then $\eta^{(1)}[\Psi,\Psi^*]$ is given as 
\begin{eqnarray}\label{app:eta^{(1)}}
\eta_i^{(1)}=\frac{\partial\Gamma^{(1)}[\Psi,\Psi^*]}{\partial\Psi_i^*}=-t\sum_{<j>}\Psi_i.
\end{eqnarray}
Up to first order in the coupling constant $g$, we have
\begin{eqnarray}\label{app:eta}
\eta_i\simeq\eta^{(0)}_i+\eta_i^{(1)}.
\end{eqnarray}
Turning off the external field, i.e., setting $\eta=0$, we find $\eta_i^{(0)}=-t\sum_{<j>}\Psi_j$. 
By substituting this expression to Eq.(\ref{app:eta^{(0)}}), we can obtain a closed equation that determines $\Psi_i$. 
We note that our first order calculation is equivalent to the mean-field theory in Ref.~\cite{oosten,buonsante}, since $\eta_i^{(0)}$ plays a role of the internal mean field obtained from the decoupling
\begin{eqnarray}
b_i^{\dag}b_j\simeq\langle b_i^{\dag}\rangle b_j+\langle b_j\rangle b_i^{\dag}=\Psi_i^*b_j+\Psi_ib_j^{\dag}.
\end{eqnarray}
The advantage of the present formalism is that one can systematically include the higher-order terms in the internal field $\eta_i^{(0)}$~\cite{georges,shiba}.

By using Eq.(\ref{d:gamma-0th}), (\ref{d:gamma-1st}) and (\ref{app:eta}), 
we finally obtain the expression for the Helmholtz free energy to 1st order in the hopping parameter $t$. 
For simplicity we present the result for the homogeneous case. 
The free energy per lattice site is given by
\begin{eqnarray}\label{d:gamma-0th+1st}
\Gamma[\Psi, \Psi^*]&=&-\frac{1}{\beta}\ln Z_0
+zt\left(1-zt\langle f\rangle\right)|\Psi|^2\nonumber\\
& &{}+
(zt)^4\biggr[\langle h\rangle+\frac{\beta}{2}\left[\langle f^2\rangle-\langle f\rangle^2\right]\biggr]
|\Psi|^4.
\end{eqnarray}
Comparing Eq.(\ref{d:gamma-0th+1st}) with Eq.(\ref{mean:landau}), we find $\Gamma^{(0)}=-\ln Z_0/\beta$ with $Z_0$ defined by (\ref{thermal-average}) and
\begin{eqnarray}
A&=&zt(1-zt\langle f\rangle),\\
B&=&(zt)^4\left\{\langle h\rangle+\beta[\langle f^2\rangle-\langle f\rangle^2]\right\},
\end{eqnarray}
with $f$ and $h$ defined by Eqs.(\ref{f}) and (\ref{h}). 
At $T=0$, these coefficients reduce to
\begin{eqnarray}
A&=&zt\left\{1-zt\left[\frac{n+1}{Un-\mu}-\frac{n}{U(n-1)-\mu}\right]\right\},\label{A}\\
B&=&(zt)^4\biggr\{
\frac{(n+1)(n+2)}{(Un-\mu)^2[U(2n+1)-2\mu]}
-\frac{(n+1)^2}{(Un-\mu)^3}\nonumber\\
&&{}-\frac{n^2}{[U(n-1)-\mu]^3}
-\frac{n(n+1)}{(Un-\mu)^2[U(n-1)-\mu]}\nonumber\\
& &{}-\frac{n(n-1)}{[U(n-1)-\mu]^2[U(2n-3)-2\mu]}\nonumber\\
&&{}
-\frac{n(n+1)}{(Un-\mu)[U(n-1)-\mu]^2}\biggr\}.
\end{eqnarray}
These results for $T=0$ agree with Ref.~\cite{oosten}.
The phase boundary between the Mott insulator (or normal) and superfluid is determined by $A=0$, and thus $zt\langle f\rangle=1$.

\section{Green's functions}\label{app:green}
The four components appearing in the expression for $G_1({\bf k},\omega)$ in Eq.(\ref{green_1}) are given by
\begin{eqnarray}
&&G_{(n)(n+1),(n+1)(n)}({\bf k},\omega)\nonumber\\
&&{}=
\frac{\hbar P_{n,n+1}\left(\hbar\omega-V_n+V_{n-1}-P_{n,n-1}T^{k}_{(n)(n-1),(n-1)(n)}\right)}
{\left(\hbar\omega-\epsilon_p({\bf k})\right)\left(\hbar\omega-\epsilon_h({\bf k})\right)}
,\label{app:green_1_1}\\
&&G_{(n)(n+1),(n)(n-1)}({\bf k},\omega)\nonumber\\
&&{}=
\frac{\hbar P_{n-1,n}P_{n+1,n}T^{k}_{(n+1)(n),(n-1)(n)}}
{\left(\hbar\omega-\epsilon_p({\bf k})\right)\left(\hbar\omega-\epsilon_h({\bf k})\right)},
\\
&&G_{(n-1)(n),(n+1)(n)}({\bf k},\omega)\nonumber\\
&&{}=
\frac{\hbar P_{n,n+1}P_{n,n-1}T^{k}_{(n)(n-1),(n)(n+1)}}
{\left(\hbar\omega-\epsilon_p({\bf k})\right)\left(\hbar\omega-\epsilon_h({\bf k})\right)},
\\
&&G_{(n-1)(n),(n)(n-1)}({\bf k},\omega)\nonumber\\
&&{}=
\frac{\hbar P_{n-1,n}\left(\hbar\omega-V_{n+1}+V_n-P_{n+1,n}T^{k}_{(n+1)(n),(n)(n+1)}\right)}
{\left(\hbar\omega-\epsilon_p({\bf k})\right)\left(\hbar\omega-\epsilon_h({\bf k})\right)},\label{app:green_1_4}
\end{eqnarray} 
while the four components appearing in the expression for $G_2({\bf k},\omega)$ in Eq.(\ref{green_2}) are given by 
\begin{eqnarray}
&&G_{(n+1)(n),(n)(n+1)}({\bf k},\omega)\nonumber\\
&&{}=
\frac{\hbar P_{n+1,n}\left(\hbar\omega-V_{n-1}+V_n-P_{n-1,n}T^{k}_{(n-1)(n),(n)(n-1)}\right)}
{\left(\hbar\omega+\epsilon_p({\bf k})\right)\left(\hbar\omega+\epsilon_h({\bf k})\right)},
\\
&&G_{(n)(n-1),(n)(n+1)}({\bf k},\omega)\nonumber\\
&&{}=
\frac{\hbar P_{n+1,n}P_{n-1,n}T^{k}_{(n-1)(n),(n+1)(n)}}
{\left(\hbar\omega+\epsilon_p({\bf k})\right)\left(\hbar\omega+\epsilon_h({\bf k})\right)},
\\
&&G_{(n+1)(n),(n-1)(n)}({\bf k},\omega)\nonumber\\
&&{}=
\frac{\hbar P_{n,n-1}P_{n,n+1}T^{k}_{(n)(n+1),(n)(n-1)}}
{\left(\hbar\omega+\epsilon_p({\bf k})\right)\left(\hbar\omega+\epsilon_h({\bf k})\right)},
\\
&&G_{(n)(n-1),(n-1)(n)}({\bf k},\omega)\nonumber\\
&&{}=
\frac{\hbar P_{n,n-1}\left(\hbar\omega-V_n+V_{n+1}-P_{n,n+1}T^{k}_{(n)(n+1),(n+1)(n)}\right)}
{\left(\hbar\omega+\epsilon_p({\bf k})\right)\left(\hbar\omega+\epsilon_h({\bf k})\right)},
\end{eqnarray}
where, $D_{\alpha}-D_{\beta}\equiv P_{\alpha,\beta}$ and $V_{\alpha,\alpha}\equiv V_{\alpha}$, and the expressions for 
$T$ and $V$ are given in Eqs.(\ref{T}) and (\ref{V}), respectively.


\begin{thebibliography}{99}
\bibitem{footnote1}In general, the optical lattice potential $V_{\rm op}'({\bf r})$ for the free state may be different from $V_{\rm op}({\bf r})$ for the trapped state. 
Moreover, the optical lattice potentials $V_{\rm op}({\bf r}')$ and $V_{\rm op}({\bf r})$ may be tuned independently by polarization or frequency selection, as proposed in Refs.~\cite{jaksch2,liu}.
\end{thebibliography}
\end{document}